\documentclass[aps,prd,twocolumn,eqsecnum]{revtex4}

\def\non{\nonumber}

\def\lsim{\begin{array}{c} < \\ \sim \end{array}}

\usepackage{amsfonts}
\usepackage{amsmath}
\usepackage{graphicx}
\usepackage{bm}
\usepackage{mathrsfs}
\usepackage{hyperref}

\begin{document}

\title{Electromagnetic and gravitational self-force on a relativistic particle from quantum fields in curved space}

\author{Chad R. Galley$^1$\footnote{crgalley@physics.umd.edu}, B. L. Hu$^{1,2}$\footnote{hub@physics.umd.edu} and Shih-Yuin Lin$^2$\footnote{sylin@phys.sinica.edu.tw} }
\affiliation{$^1$Department of Physics, University of Maryland,
College Park, Maryland, 20742 }
\affiliation{$^2$Center for Quantum and Gravitational Physics, Institute of Physics, Academia Sinica, Nankang, Taipei 11529, Taiwan}


\begin{abstract}
	We provide a quantum field theoretical derivation of the Abraham-Lorentz-Dirac (ALD) equation, describing the motion of an electric point charge sourcing an electromagnetic field, which back-reacts on the charge as a self-force, and the Mino-Sasaki-Tanaka-Quinn-Wald (MSTQW) equation describing the motion of a point mass with self-force interacting with the linearized metric perturbations caused by the mass off an otherwise vacuous curved background spacetime. We regularize the formally divergent self-force by smearing the direct part of the retarded Green's function and using a quasilocal expansion. We also derive the ALD-Langevin and the MSTQW-Langevin equations with a classical stochastic force accounting for the effect of the quantum fluctuations in the field, which causes small fluctuations on the particle trajectory. These equations will be useful for studying the stochastic motion of charges and small masses under the influence of both quantum and classical noise sources, derived either self-consistently or put in by hand phenomenologically. We also show that history-dependent noise-induced drift motions could arise from such stochastic sources on the trajectory that 
could be a hidden feature 
of gravitational wave forms hitherto unknown.
\end{abstract}

\maketitle

\section{Introduction}

Recent years have seen rapidly growing interest in the issues and problems of radiation reaction on particle motion in a curved spacetime as a result of the need to produce accurate gravitational waveforms for the proposed space-based gravitational interferometer LISA. For generic particle-field interactions the back reaction of emitted radiation on the particle results from two types of interactions with the radiation. The first is a purely local effect and describes the force on the particle when it emits radiation. The second is non-local in nature resulting from previously emitted radiation back-scattering off the background spacetime curvature and interacting with the particle at a later time in some other position. The effects of the emitted radiation appear as a force on the particle, the  self-force, that changes the particle's motion from a simple geodesic in the background spacetime. DeWitt and Brehme \cite{DeWitt_Brehme} were the first to study the electromagnetic self-force in a curved spacetime. The corresponding equation depicting the dynamics of a point mass experiencing self-force from interactions with the (linear) metric perturbations it generates off a vacuum background spacetime was first obtained  by Mino, Sasaki and Tanaka \cite{MST} and later confirmed by Quinn and Wald \cite{Quinn_Wald} using axiomatic methods. For an excellent review of these derivations and other issues concerning the motion of particles in a curved spacetime see \cite{Poisson}.

Meanwhile, the introduction and development of world line path integral techniques (see \cite{JH1} and references therein) allows one to derive the equations of motion of (relativistic) particles interacting with quantum fields within a self-consistent formalism. Augmented by the paradigm of open quantum systems provides one with a self-consistent means for studying the effects of quantum field fluctuations on the particle's motion. One can show {\it ab initio} how these fluctuations manifest as classical stochastic forces (see \cite{RHA, RHK, RavalPhD} for the case of accelerating detectors (atoms) and \cite{JH1, JHIARD, JohnsonPhD} for moving charges). This approach can be applied to the relation between Unruh radiation and ordinary radiation in accelerating detectors (see \cite{HRCapri, LH1, RHA}).


In a previous paper \cite{GH1} we use the world line influence functional approach to obtain the equations of motion for a scalar charge coupled to a quantum scalar field in a curved spacetime.
Our results for the classical motion agree with Quinn \cite{Quinn}. We also derive a scalar ALD-Langevin equation and show that the fluctuations in the trajectory depend on their past history (non-Markovian), implying that secular effects may be important when stochastic fluctuations are included in a more complete description of the particle's motion. 


In this paper, we use the same approach to study the motions of two kinds of objects. The first is the dynamics of an electric point charge moving in a quantum electromagnetic field in a curved spacetime. This generalizes results obtained earlier \cite{JH1} from motion in a flat spacetime to a curved one. The second is the dynamics of a small mass (point particle) 
interacting with quantized linear metric perturbations off a background spacetime (e.g. a massive black hole). In both cases we are interested in the radiation the particle emits and how its back reaction alters the trajectory. Our aim is to derive the equations of motion for the quantum expectation value of the particle's position. If the mechanisms that decohere the particle is strong enough then the expectation value behaves classically. In this limit we regain the well-known Abraham-Lorentz-Dirac (ALD) equation in a curved spacetime, which agrees with the earlier derivations of DeWitt and Brehme \cite{DeWitt_Brehme}, and the Mino-Sasaki-Tanaka-Quinn-Wald (MSTQW) equation, respectively \cite{MST, Quinn_Wald}.

We observe that the self-force acting on the particle is formally infinite as a result of considering point-particle interactions with the field. This is the usual ultraviolet divergence arising from the assumption of a point particle. It must be regularized to obtain a finite and unambiguous self-force. We use a method introduced and developed in \cite{JH1, GH1} that is inspired by effective field theory, which provides a natural way to quickly identify and renormalize this divergence. A high energy scale is introduced into the dynamics so that 
certain criteria on the magnitude of this scale and others must be met in order to have a dynamics consistent with our first principles derivations of (semi-classical) equations. This regularization procedure is not restricted to quantum fields so that one can apply this to the usual problem of relativistic particle and classical field dynamics. 

After the derivation of the semi-classical (noise-averaged) equations of motion we incorporate the effect of quantum field fluctuations on the motion of the particles by showing how they behave like classical stochastic forces. We derive an ALD-Langevin equation for the electric charge motion and a MSTQW-Langevin equation for the point mass motion. These equations describe the actual trajectory including stochastic components arising from the quantum field fluctuations when the world line histories are sufficiently decohered. They are useful for studying the stochastic motion of charges and point masses under the influence of quantum noise sources in curved spacetime. The same form of the Langevin equations derived here with quantum field-induced noise can be used for classical stochastic forces in a phenomenological description of astrophysical sources in stochastic motion. 

For such classical stochastic forces we show that the correlations of the stochastic noise may give rise to a noise-induced drift force that is analogous to the particle drifts encountered in plasma physics. For noise derived from the quantum field fluctuations this effect should contain some information about the quantum statistical state of the particle \cite{Calzetta_Roura_Verdaguer}. Our study makes it possible to explore in more detail these novel phenomena that may have some observational effects in certain atomic optics experiments. This drift might also intercede in obtaining more accurate calculations of gravitational waveform templates, although there are likely no significant quantum fluctuation-induced stochastic effects in astrophysical sources.

This paper is organized as follows. In Section II, after a brief summary of the world line influence functional method we describe how to obtain the semiclassical and stochastic semiclassical particle dynamics for a relativistic particle interacting with a tensor field. We also describe the philosophy and implementation of the effective field theory inspired approach for regularizing the self-force. In Section III we apply our results to the specific example of a point charge moving through a quantum vector field in a curved spacetime. In Section IV we study the motion of a point mass moving in a curved background spacetime with quantum linear metric perturbations. In Section V we discuss a phenomenological description of stochastic particle dynamics by adding in noise by hand. This added noise is shown to generate a noise-induced drift in the particle motion that is analogous to the usual plasma drifts. In Section VI we discuss the various scales appearing in our approximations in order to obtain well-defined solutions that are free from the usual pathologies (e.g. runaway solutions, pre-acceleration, etc.). Section VII contains a summary of our results and Appendices A, B and C provide further details of certain calculations used in the text.

Most of the treatment here parallels that in \cite{GH1}, which the reader is referred to for more details. We use units where $c=G=1$, unless otherwise stated, and use the conventions of Misner, Thorne, and Wheeler \cite{MTW} with signature $(-,+,+,+)$. We also use the notation that an unprimed (primed) index refers to that component of a tensor field or coordinate evaluated at the point $x$ ($x^\prime$) or proper time $\tau$ ($\tau^\prime$), as appropriate (e.g., $z^{\alpha^\prime} = z^\alpha (\tau^\prime)$).

\section{Relativistic Particle-Field Dynamics}

We choose the world line coordinates $z^\mu (\tau)$ as the system of interest interacting with a quantum field $\Phi_A (x)$ considered as its environment. The subscript $A$ denotes possible tensor indices. We are only interested in the overall influence, but not the detailed behavior of the environment so we will coarse-grain over the quantum field variables. For this purpose the world line influence functional is used here as in \cite{JH1} to determine the influence of the field on the particle's motion in a curved background spacetime (with metric $g_{\mu\nu}$).


\subsection{World line influence functional}


Assume at some initial time $t_i$, for a given coordinate system, that the quantum statistical state of the system (particle) and environment (quantum field) is described by a density matrix $\rho (z_i, \Phi_{Ai} ; z_i ^\prime , \Phi_{Ai}^\prime ; t_i)$. For purely technical convenience it is customary to choose the initial density matrix to correspond to a factorized state of the system and environment
\begin{eqnarray}
&& \!\!\!\! \rho (z_i, \Phi_{Ai}; z_i ^\prime, \Phi_{Ai} ^\prime; t_i ) = \rho_S
(z_i, z_i ^\prime; t_i ) \otimes \rho_E (\Phi_{Ai}, \Phi_{Ai} ^\prime;
t_i ) ~. \non 
\end{eqnarray}
Physically, this means that all of the field modes have been uncorrelated with the particle at time $t_i$. We will also assume that the initial state of the environment $\hat{\rho}_E$ is Gaussian in the initial field configurations. (In practice, there may be radiation present prior to the measurement at the initial time.)
Throughout this paper we will explicitly retain the dependence on the initial time.

We take the action for the complete (closed) system to consist of an action describing the free evolutions of the system of interest, that of a relativistic point particle of mass $m_0$, the environment, that of a quantum field (to be coarse-grained), and their mutual interaction, $S_{S+E} [z,\Phi_A] = S_S [z] + S_E [\Phi_A] + S_{int} [z,\Phi_A]$. The system action is given by
\begin{eqnarray}
	S_S [z] &=& - m_0 \int d\tau \sqrt{-g_{\mu\nu} u^\mu u^\nu } 
	\label{sys_action}
\end{eqnarray}
where $u^\mu$ is the particle's 4-velocity and $\tau$ is the world line's parameter (not necessarily the proper time at this stage). The action for the linear quantum field is given by
\begin{eqnarray}
	S_E [\Phi_A] = \frac{1}{2} \int d^4x \, \sqrt{-g} \, \Phi_A \vec{\cal D} ^{AB} \Phi_B
\end{eqnarray}
where $\vec{\cal D} _{AB}$ is the kinetic operator appropriate to the quantum field. The interaction action is linear in the field variable
\begin{eqnarray}
	S_{int} [z,\Phi_A] &=& \int d^4 x \, \sqrt{-g} \, j^A (x;z) \Phi_A (x) 
	\label{int_action}
\end{eqnarray}
where the current density $j_A (x;z)$ is some functional of the world line coordinates whose specific form will depend upon the field chosen. 

Since the quantum field may have gauge degrees of freedom it is necessary to include a gauge-fixing term to the action. If the gauge choice is implemented through the constraint $G_B(\Phi_A) \approx 0$ for some appropriately chosen function $G_B$ (the $\approx$ denotes weak equality in the sense of Dirac \cite{Dirac}) then the procedure of Faddeev and Popov \cite{Faddeev_Popov} amounts to introducing the following gauge-fixing term to the field action
\begin{eqnarray}
	S_{gf} = - \alpha \int d^4 x \, \sqrt{-g} \, G_B G^B 
\end{eqnarray}
where $\alpha$ is some constant that can be chosen rather arbitrarily. In this paper we will be dealing with tree-level fields exclusively so there is no need to keep track of the ghost fields in the action. For derivations in the remainder of this section we assume that the function $G_B$ is approximately linear in the field so that the gauge-fixing action $S_{gf}$ is quadratic. Any nonlinear term that might appear in $G_B$ we assume to be small and negligible within the context of the approximations used below.

We coarse-grain the field variables by tracing them out from the time-evolved density matrix to get the {\it reduced density matrix} for the system
\begin{eqnarray}
    && \rho_r (z_f, z_f ^\prime; t_f ) = \int dz_i \, dz_i ^\prime \int _{z_i} ^{z_f} {\cal D}z \int _{z_i ^\prime} ^{z_f ^\prime} {\cal D} z^\prime \, \rho _S (z_i, z_i ^\prime; t_i ) \non \\
        && {\hskip 0.5in} \times   \, e^{ i / \hbar ( S_S [z] - S_S [z ^\prime] ) } \, F[z, z^\prime]  
        \label{red_rho}
\end{eqnarray}
for times $t_f > t_i$. The quantity $F[z,z^\prime]$ is the {\it influence functional}
\begin{eqnarray}
	F[z,z^\prime] = e^{-(1/4\hbar) \, j^- _A \cdot G_H ^{AB^\prime} \cdot j^- _{B^\prime} + (i/\hbar) \, j^- _A \cdot G_{ret} ^{AB^\prime} \cdot j^+ _{B^\prime} }
	\label{IF}
\end{eqnarray}
which describes the influence of the coarse-grained environment on the particle histories $z$ and $z^\prime$. (The $\cdot$ denotes spacetime integration between times $t_i$ and $t_f$ and the difference $j^-_A$ and semi-sum $j^+_A$ current densities are defined as the difference and average of $j_A(x;z)$ and $j_A (x; z^\prime)$, respectively.) The Hadamard $G ^{AB^\prime}_H$ and retarded Green's functions $G^{AB^\prime}_{ret}$ are given by
\begin{eqnarray}
	G^H_{AB^\prime} ( x, x^\prime ) &=& \big\langle \{ \hat{\Phi}_{AI} (x), \hat{\Phi}_{B^\prime I} (x^\prime) \} \big\rangle \non \\
    	G^{ret} _{AB^\prime} (x, x^\prime ) &=& i \, \theta_+ ( x, \Sigma ) \: \big\langle [ \hat{\Phi}_{AI} (x), \hat{\Phi}_{B^\prime I} (x^\prime) ] \big\rangle  ~. \non \\
       \label{ret}
\end{eqnarray}
and satisfy $\vec{\cal D}^{AB} G^H _{BC^\prime} = 0$ and $\vec{\cal D}^{AB} G^{ret} _{BC^\prime} = g^A _{~C^\prime}$, respectively, where $g^A_{~C^\prime}$ symbolically represents the delta-function source appropriate for the quantum field equations of motion. The above are quantum fields in the interaction picture $\Phi_{AI}$ and the quantum expectation values are given by $\langle \cdots \rangle = {\rm Tr}_E \, \hat{\rho}_E (\cdots)$ with respect to the Gaussian initial state $\hat{\rho}_E$ of the environment. The step function $\theta_+ (x, \Sigma)$ appearing in $G ^{AB^\prime}_{ret}$ equals one
in the future of the point $x^\prime$ and zero otherwise, and $\Sigma$ is a space-like hypersurface containing $x^\prime$. 


\subsection{Semiclassical particle dynamics}
\label{semiclassical_section}

The reduced density matrix (\ref{red_rho}) for the particle can now be written as
\begin{eqnarray}
  && \!\!\!\!\!\! \rho_r (z_f, z_f ^\prime; t_f ) \non \\
    && = \!\!\!\! \int dz_i \, dz_i ^\prime \int _{z_i} ^{z_f} \!\! {\cal D} z \int _{z_i^\prime} ^{z_f ^\prime} \!\! {\cal D} z^\prime \, \rho_S (z_i, z_i^\prime; t_i ) \, e^{ ( i / \hbar ) S_{CGEA} [z, z^\prime] } \non \\
\end{eqnarray}
where the {\it coarse-grained effective action} (CGEA) is defined as
\begin{eqnarray}
   && \!\!\!\!\!\! S_{CGEA} [z, z^\prime]  \non \\
        && = S_S [z]  - S_S [z^\prime] + j^-_A \cdot  G^{AB^\prime}_{ret}  \cdot j^+ _{B^\prime} + \frac{i}{4} \, j^-_A \cdot  G^{AB^\prime}_{H} \cdot j^-_{B^\prime} \non \\
        && \label{cgea}
\end{eqnarray}
and provides a description of the quantum-averaged particle dynamics. The magnitude of the influence functional, which is a measure of decoherence \cite{conhis}, decays exponentially fast for two largely separated particle histories as a result of the environmentally induced decoherence of the particle through its interactions with the quantum field fluctuations
\begin{eqnarray}
	| F | &=& e^{ -(1/4 \hbar) j^- _A \cdot G_H ^{AB^\prime} \cdot j^- _{B^\prime} } \non \\
	&=& e^{-(1/4 \hbar) \int d\tau d\tau^\prime z_- ^\alpha z_- ^{\beta^\prime} \vec{w}_\alpha ^{~A} \vec{w}_{~\beta^\prime} ^{B^\prime} G^H _{A B ^\prime} + O(z_- ^4) }  ~.
\end{eqnarray}
 If these fluctuations provide a strong enough mechanism of decoherence, which depends on the magnitude of $G_H ^{AB^\prime}$, then we are justified in expanding the CGEA about the classical worldline $\bar{z}^\mu (\tau)$ for which the small difference of the histories $z^\mu _- = z^\mu - z^{\mu \prime}$, resulting from quantum fluctuations of the particle itself, gives the dominant contribution to the path integral in the reduced density matrix. 
A stationary phase approximation of the resulting reduced density matrix gives the equations of motion for the classical worldline $\bar{z}^\mu$
\begin{eqnarray}
	\left. \frac{ \delta S_{CGEA} }{\delta z^\mu _- (\tau) } \right|_{z = z^\prime = \bar{z} } {\hskip-0.2in} = 0
	\label{cgea_eom}
\end{eqnarray}
which, in terms of the actions (\ref{sys_action}) and (\ref{int_action}), are
\begin{eqnarray}
	m_o \bar{a}_\mu = \vec{w}_\mu ^{~A } [\bar{z}] \, \Phi _{A} ^{ret} (\bar{z})  ~.
	\label{cl_eom}
\end{eqnarray}
The proper time gauge $\bar{u}_\mu \bar{u}^\mu = -1$ has been chosen. Here, $\vec{w}_\mu ^{~A}$ is a derivative operator obtained from the functional derivative of the current density $j^A (x;z)$ and $\Phi_{A} ^{ret}$ is the retarded propagator of the quantum field evaluated in the initial state $\hat{\rho}_E$. The gauge degrees of freedom of the field $\Phi_A^{ret}$ will (usually) be projected out by $\vec{w}_\mu ^{~A }$. The acceleration on the left side is $D \bar{u}_\mu / d\tau$ where $D/d\tau = \bar{u}^\beta \nabla _\beta$. Throughout, we use an overbar to denote classical quantities.

\subsection{Stochastic semiclassical particle dynamics}
\label{stoch_formalism}

Even under the assumption of strong decoherence, when the classical
trajectory becomes well-defined, the quantum fluctuations of the field
can still influence the classical motion of the particle
through the particle-field coupling $S_{int}$ and manifest as
classical stochastic forces on the particle. We show how this comes about using the influence functional formalism.

We start by invoking the relation {\it a la} Feynman and Vernon \cite{Feynman_Vernon},
\begin{eqnarray}
    && e^{- (1 / 4 \hbar ) j^-_A \cdot G^{AB^\prime}_H \cdot j^-_{B^\prime}}  \non \\
    && {\hskip0.25in} = N \!\! \int {\cal D} \xi(x) \, e^{- ( 1 / \hbar ) \xi_A \cdot ( G^{AB^\prime}_H )^{\, -1} \cdot \xi_{B^\prime} - ( i / \hbar ) \xi^A \cdot j^-_A} \non 
\end{eqnarray}
where $N$ is a normalization factor that is independent of the worldline coordinates and $\xi_A(x)$ is some auxiliary field. Now the reduced density matrix (\ref{red_rho}) becomes
\begin{eqnarray}
    \rho_r (z_f, z_f ^\prime; t_f) \!\! &=& \!\! N \!\! \int dz_i \, dz_i ^\prime \int _{z_i} ^{z_f} \!\! {\cal D} z \int_{z_i ^\prime} ^{z_f^\prime} \!\! {\cal D} z^\prime \, \rho_S (z_i, z_i ^\prime; t_i ) \non \\
    && \times \int {\cal D} \xi_A \, P_\xi [\xi_A] \, e^{ i/\hbar S_{SEA} [z, z^\prime; \xi_A] }
    \label{rho_sea}
\end{eqnarray}
where the {\it stochastic effective action} (SEA) is defined as
\begin{eqnarray}
    S_{SEA} [z, z^\prime; \xi_A] = \Re S_{CGEA}  [z,z^\prime] -  \xi^A \cdot j^-_A  ~.
    \label{SEA}
\end{eqnarray}
The function $\xi_A(x)$ can be
interpreted as a classical stochastic, or noise, field \cite{JH1, Calzetta_Roura_Verdaguer}
with an associated (Gaussian) probability distribution functional
\begin{eqnarray}
P_\xi [\xi_A(x)] = e^{- ( 1 / \hbar ) \xi_A \cdot ( G^{AB^\prime}_H ) ^{\, -1} \cdot \xi_{B^\prime} }  ~.
\end{eqnarray}
The fact that this is Gaussian is a direct consequence of the quadratic field action $S_E[\Phi]$ and the field's linear coupling to the particle current density in $S_{int}[z,\Phi]$.
With respect to $P_\xi [\xi_A]$ this implies that $\xi_A$ has zero-mean
and its correlator is proportional to the Hadamard function
encoding the information about the fluctuations in the quantum
field \cite{footnote_1}
\begin{eqnarray}
    \big\langle \{ \xi^A (x), \xi^{B^\prime} (x^\prime) \} \big\rangle _\xi &=& \hbar \, G^{AB^\prime}_H (x, x^\prime)   \label{xi_correlator}
\end{eqnarray}
where $\langle \ldots \rangle _\xi = N \int {\cal D} \xi_A \, P_\xi \, (\ldots)$.

Expanding the SEA around the classical solution $\bar{z}^\mu$ and evaluating the resulting reduced density matrix using the stationary phase approximation results in the following stochastic equations of motion for the world line fluctuations $\tilde{z}^\mu \equiv z^\mu - \bar{z}^\mu$
\begin{eqnarray}
	\left. \int d\tau^\prime \, \tilde{z}^{\alpha^\prime} \frac{ \delta^2 \Re S_{CGEA} }{ \delta z ^{\alpha^\prime} \delta z  ^\mu } \right|_{z=z^\prime = \bar{z}} \!\!\! = \eta_\mu [\bar{z}]
	\label{sea_eom}
\end{eqnarray}
where the stochastic force is 
\begin{eqnarray}
    \eta_\mu [z] = \vec{w}_\mu ^{~A}[z] \, \xi_A(z)  ~.
\end{eqnarray}
This equation describes the dynamics of small perturbations $\tilde{z}$ around the semiclassical solution $\bar{z}$ that originate from the classical, stochastic manifestation $\eta_\mu$ of the quantum field fluctuations. We can obtain a stochastic version of (\ref{cl_eom}) if we add (\ref{cgea_eom}) to the left side of (\ref{sea_eom}), i.e., 
\begin{eqnarray}
	\left. \frac{ \delta \Re S_{CGEA} }{ \delta z^\mu _-} \right| _{z^- = 0}  = \eta_\mu [z]  ~.
\end{eqnarray}
The functional derivative gives the final form of the stochastic equations of motion in terms of the full world line $z^\mu$
\begin{eqnarray}
    m_o a_\mu = \vec{w}_\mu ^{~A} [z] \, \Phi_{A}^{ret} (z) + \eta_\mu [z]  ~.
\label{stoch_eom}
\end{eqnarray}
This equation is only valid to linear order in the fluctuations $\tilde{z}$, which is the same order as the stochastic force $\eta_\mu[\bar{z}]$, because we are neglecting higher order quantum corrections. In practice, (\ref{stoch_eom}) is expanded to linear order in $\tilde{z}$ and (\ref{cl_eom}) is invoked to obtain the particle fluctuation dynamics. We point out that (\ref{stoch_eom}) is a stochastic equation of motion because observables involving $\tilde{z}$ must be computed via the stochastic correlation functions $\langle \ldots \rangle_\xi$. Also, notice that both the deterministic and the stochastic components of the self-force can push the particle away from its mean trajectory with respect to a fixed background spacetime. 

The stochastic correlation functions of the force $\eta_\mu$ can
be evaluated using the $\xi_A$ correlators above.
Evaluating these correlation functions along the classical trajectory $\bar{z}$ we find that the mean of the stochastic force is zero and the symmetric two-point function of the stochastic force (noise kernel) is
\begin{eqnarray}
    \big\langle \{ \eta_\mu [ \bar{z} ], \eta_{\nu^\prime} [ \bar{z} ] \} \big\rangle _\xi =  \hbar \, \vec{w}_{ ( \mu  } ^{~~A} [\bar{z}] \, \vec{w}_{\nu^\prime ) } ^{~~B^\prime} [\bar{z}] \, G^H_{AB^\prime } (\bar{z}^\alpha, \bar{z}^{\alpha^\prime} )  ~. \non \\
\label{eta_corr}
\end{eqnarray}
It follows that the noise $\eta_\mu$, which is generally multiplicative and colored, depends on the particle's initial conditions through the classical trajectory and on the field's initial conditions via $G^{AB^\prime}_H$. For most fields (with the notable exception of the linearized metric perturbation $h_{\mu\nu}$) the projection operator $\vec{w}_\mu ^{~A}$ enforces the noise kernel to be gauge invariant.
For equal proper times $\tau^\prime = \tau$ the Hadamard function diverges so a suitable regularization procedure
must be used in order to make sense of (\ref{eta_corr}) near coincidence.

The noise kernel (\ref{eta_corr}) shows that the stochastic force $\eta_\mu$ is $O(\hbar^{1/2})$ as are the world line fluctuations $\tilde{z}$. This shows that the Langevin equation (\ref{sea_eom}) is between the tree-level 
 and the one-loop $O(\hbar)$ equations of motion and therefore contains information about the lowest order \footnote{Lowest order in the coupling constant, that is. In this paper we consider a linear quantum field (or more generally, a quantum field in the Gaussian approximation), which has no non-trivial vertices and hence no loop contributions to the field two-point functions, e.g. $G_H$ and $G_{ret}$.}
quantum fluctuations of the coarse-grained environment. This is true even if the environment is weakly nonlinear. This is the reason why we do not need to include the ghost fields in these computations since the ghosts first appear at one-loop order and hence provide no contribution to the stochastic semi-classical motion of the particle.

\subsection{Retarded Green's function}

In order to study the self-force we will need to know the form of the retarded Green's function used to construct the retarded field $\Phi ^{ret}_A (z)$. Assuming strong environment-induced decoherence it follows that the field is sourced by a classical particle current $j(\bar{z})$
allowing us to invoke Hadamard's ansatz, which takes the form in a curved background spacetime
\begin{eqnarray}
	G^{ret} _{A B^\prime} (x,x^\prime) &=& U_{A B^\prime}(x,x^\prime) \Delta^{1/2} (x,x^\prime) \delta_+ (\sigma(x,x^\prime)) \nonumber \\
	&& + V_{A B^\prime} (x,x^\prime) \theta_+ (-\sigma(x,x^\prime) )
\end{eqnarray}
as the sum of a ``direct" part (proportional to $\delta_+$) and a ``tail" part (proportional to $\theta_+$). This ansatz assumes that $x$ and $x^\prime$ are connected by a unique geodesic since otherwise the van Vleck determinant $\Delta (x,x^\prime)$ will diverge at a caustic. 
The tensor $U_{AB^\prime}$ involves products of the bi-tensor $g_\alpha ^{~\beta^\prime} (x,x^\prime)$ that parallel transports a tensor at $x^\prime$ to $x$. 
The tensor $V_{AB^\prime}$ satisfies the homogeneous field equations subject to characteristic data provided along the forward lightcone at $x^\prime$ (see \cite{Poisson} for details). This tensor field is quite difficult to calculate for generic spacetimes so throughout the remainder we consider this object only formally. 

The distributions $\delta_+$ and $\theta_+$ are defined as $\theta_+ (x,\Sigma)$ multiplying $\delta (\sigma)$ and $\theta (-\sigma)$, respectively, where $\sigma = \sigma (x, x^\prime)$ is Synge's world function along the (unique) geodesic linking  $x=z(\tau)$ and $x^\prime=z(\tau^\prime)$ and is half of the squared geodesic distance between the two points. 

\subsection{Regularization motivated by effective field theory}
\label{sec_eft_reg} 

The presence of the $\delta(\sigma)$ term in the direct part of the retarded Green's function indicates a divergence in the retarded field $\Phi^{ret}_A = G^{ret}_{AB} \cdot j^B$ when the two points $x$ and $x^\prime$ coincide under the integral. This is the usual divergence that results from considering interactions between a point-particle and arbitrarily high modes of a quantum field. We use the regularization introduced in \cite{JH1} (in flat spacetime) and extended to scalar particle dynamics in curved spacetimes in \cite{GH1}. 


In this spirit we introduce a regulator $\Lambda$ for the field so that for energies much lower than the associated energy scale $E_\Lambda$ we expect the semiclassical (\ref{cl_eom}) and stochastic semiclassical (\ref{stoch_eom}) dynamics to accurately describe the particle motion.
This approach is taken in \cite{JH1} for deriving the ALD equations in flat spacetime and is easily extended to motions in a curved spacetime \cite{GH1} since the ultraviolet divergence is local in origin.

To regularize the divergence we make the replacement
\begin{eqnarray}
	\delta_+ (\sigma(x^\alpha, x^{\prime \alpha} ) ) &\rightarrow& \theta_+ (-\sigma) \delta_\Lambda (\sigma) \non \\
	&& = \theta_+ (x, \Sigma) \theta (-\sigma) \sqrt{ \frac{8}{\pi} } \, \Lambda^2 e^{- 2 \Lambda^4 \sigma^2 } \non \\
	\label{reg}
\end{eqnarray}
in the direct part of the retarded Green's function (or retarded field $\Phi_A ^{ret}$). 
The function $\delta_\Lambda (\sigma)$ is smooth but approximates $\delta (\sigma)$ well only if $\Lambda^2 \sigma \gg 1$. This inequality will not hold if $\sigma$ is strictly zero so we will assume that $\sigma$ is small and approaching zero while maintaining $\Lambda^2$ significantly larger than $1/\sigma$. For time-like separated points, e.g., points on a particle trajectory, $\theta (-\sigma) = 1$. 

Since $\Lambda$ serves to provide a minimum resolution the sharp step function $\theta_+ (x, \Sigma)$ should be replaced by the smooth function
\begin{eqnarray}
	\theta_+ (x,\Sigma) \rightarrow \theta_\Lambda (-s) = \frac{ 2^{1/4} \sqrt{\pi} }{ \Gamma(1/4) } \int _{-\infty} ^{-s} ds^\prime \, \delta_\Lambda (\sigma (s) ) \non
\end{eqnarray}
when $x$ and $x^\prime$ are on a particle world line. For such a pair of points the dominant contribution to $\delta_+$ comes from those points that are nearly coincident so that the proper time difference $s = \tau^\prime - \tau$ is small but still much larger than the resolution scale $1/\Lambda$. Using the results from Appendix A it is easy to show that for large $\Lambda$
\begin{eqnarray}
	\theta_\Lambda (-s) \sim \theta (-s) + \frac{1}{2^{1/4} \Gamma(1/4) } \, \frac{  {\rm sgn}(s) }{s^3 \Lambda^3} \, e^{-\Lambda^4 s^4/2} \non
\end{eqnarray}
The last term falls off faster than any inverse power of $s \Lambda$, which allows us to ignore the smooth effects of the regulator on the step function so that we only need to keep the first term $\theta(-s)$ in all that follows.

In the next sections we study the two specific examples of a point charge moving through a quantum vector field $A_\mu (x)$ and a point (infinitesimal) mass moving through a quantum field of (linearized) metric perturbations, both on a given background spacetime.

\section{Electromagnetic Field}
\label{em}

In this section we will study the specific example of a point charge $e$ moving through a quantum vector field with $\Phi_A (x) = A_\mu (x)$ in a curved background spacetime. The actions describing the dynamics of the field (environment) and its interaction with the particle (system) are
\begin{eqnarray}
	S_E [A_\mu] &=& - \frac{1}{4} \int d^4x \, \sqrt{-g} \, F_{\mu\nu} F^{\mu\nu} \non \\
	S_{int} [z, A_\mu] &=& \int d^4x \, \sqrt{-g} \, j^\mu (x;z) A_\mu (x)
\end{eqnarray}
where the current density is 
\begin{eqnarray}
	j^\mu (x; z) = e \int d\tau \, \frac{ \delta ^4 (x-z) }{ \sqrt{-g} } \, g^\mu _{~\alpha } (x,z) \, u^\alpha
\end{eqnarray}
and $F_{\mu\nu} = \nabla_\mu A_\nu - \nabla _\nu A_\mu$. We choose the Lorentz gauge so that the gauge-fixing action is
\begin{eqnarray}
	S_{gf} [A_\mu] = - \frac{1}{ 2} \, \int d^4x \, \sqrt{-g} \, G^2
\end{eqnarray}
where $G=\nabla_\mu A^\mu$ is the gauge-fixing function. From these we can construct the influence functional
\begin{eqnarray}
	F[z,z^\prime] = e^{-(1/4\hbar) j_\mu ^- \cdot G_H ^{\mu\nu^\prime} \cdot j_{\nu^\prime} ^- + (i /\hbar) j_\mu^- \cdot G_{ret}^{\mu\nu^\prime} \cdot j^+_{\nu^\prime} }
	\label{em_inf}
\end{eqnarray}
where 
\begin{eqnarray}
	&& G_H ^{\mu\nu^\prime} = \big\langle \{ \hat{A}^\mu , \hat{A}^{\nu^\prime} \} \big\rangle \non \\
	&& G_{ret} ^{\mu\nu^\prime} = i \, \theta_+ (x, \Sigma) \, \big\langle [ \hat{A}^\mu, \hat{A}^{\nu^\prime} ] \big\rangle
\end{eqnarray}
and obtain the motion of the particle at the semiclassical and stochastic semiclassical levels.

\subsection{Semiclassical dynamics}

Using (\ref{em_inf}) we compute the coarse-grained effective action and find that
\begin{eqnarray}
	&& \!\!\!\!  S_{CGEA} [z,z^\prime] \nonumber \\
	&&  = S_S [z]z - S_S[z^\prime] + j_\mu ^- \cdot G_{ret} ^{\mu\nu^\prime} \cdot j_{\nu^\prime} ^+ + \frac{i}{4} \, j_\mu ^- \cdot G_H ^{\mu\nu^\prime} \cdot j_{\nu^\prime} ^-  ~. \nonumber \\
\end{eqnarray}
From the discussion in Section \ref{semiclassical_section}, varying this action yields the semiclassical equation of motion for the particle (\ref{cgea_eom}) 
\begin{eqnarray}
	m_o \bar{a}_\mu = e \vec{w}_\mu ^{~\alpha} [\bar{z}] A_\alpha ^{ret} (\bar{z}) 
	\label{em_cl_eom}
\end{eqnarray}
where the derivative operator $\vec{w} _\mu ^{~\alpha}$ is
\begin{eqnarray}
	\vec{w}_\mu ^{~\alpha} [z] = - w_\mu ^{~\alpha \beta} [z] \nabla_\beta = - 2 g_\mu ^{~[\alpha} (z) \, u^{\beta]} \nabla_\beta ~.
\end{eqnarray}
It follows from the functional derivative of the current density $j^\alpha$ integrated over a suitable test function
\begin{eqnarray}
	\frac{\delta}{ \delta z^\mu (\tau) } \int d^4 x \, \sqrt{-g} \, j^\alpha (x ; z ) f(x) = e \, \vec{w}_\mu ^{~\alpha} [z] \, f(z)  ~.
\end{eqnarray}

The retarded vector field is expressed in terms of the retarded Green's function $A^{ret}_\alpha = G^{ret} _{\alpha \beta^\prime} \cdot j^{\beta^\prime}$. Using Hadamard's ansatz for the two points $x = z(\tau)$ and $x^\prime = z(\tau^\prime)$ connected by a unique geodesic, the retarded Green's function is
\begin{eqnarray}
	G^{ret} _{\alpha \beta^\prime} (x,x^\prime) &=& g_{\alpha \beta^\prime} (x,x^\prime) \Delta^{1/2} (x,x^\prime) \delta_+ (\sigma(x,x^\prime) \nonumber \\
	&& + V_{\alpha \beta^\prime} (x,x^\prime) \theta_+ (-\sigma (x,x^\prime) )
\end{eqnarray}
where $g_{\alpha\beta^\prime}$ is the bi-tensor of parallel transport and $V_{\alpha \beta^\prime}$ satisfies the homogeneous field equation
\begin{eqnarray}
	\Box V_{\alpha \beta^\prime} - R_\alpha ^{~\mu} V_{\mu\beta^\prime} = 0
\end{eqnarray}
subject to certain characteristic data determined on the forward lightcone at $x^\prime$ (see \cite{Poisson} for details). Note that the direct part of the Green's function gives rise to a divergence when the two points are light-like separated. In particular, the self-force in (\ref{em_cl_eom}) requires its evaluation along the world line so that the only contribution coming from the direct part arises when the two points are equal. In the next section we regularize this divergence using the effective field theory approach.

\subsection{Regularization}

Introduce a regulator $\Lambda$ such that for energies lower than the scale of this regulator $E_\Lambda$ an effective point-particle description of the particle dynamics becomes suitable. Making the replacement (\ref{reg}) in the direct part of the retarded Green's function we find that (\ref{em_cl_eom}) becomes, after passing the covariant derivative through the integral,
\begin{eqnarray}
	m_o \bar{a}_\mu \!\! &=& \!\! - e^2 w_\mu ^{~\alpha \beta} \! \left[ (\tau - \tau^\prime)_{;\beta} \right] \! \left[ g_{\alpha \gamma^\prime} u^{\gamma^\prime} \Delta^{1/2} \delta_\Lambda + V_{\alpha \gamma^\prime} u^{\gamma^\prime} \right] \nonumber \\
	&& - e^2 w_\mu ^{~\alpha \beta} \int _{\tau_i} ^\tau d\tau^\prime \left\{ \delta_\Lambda (\sigma) \nabla_\beta \left( g_{\alpha \gamma^\prime} u^{\gamma^\prime} \Delta^{1/2} \right) \right. \nonumber \\
	&& \left. + g_{\alpha \gamma^\prime} u^{\gamma^\prime} \Delta^{1/2} \nabla_\beta \delta_\Lambda ( \sigma) \right\} + e w_\mu ^{~\alpha \beta} A_{\alpha \beta} ^{tail} (\bar{z})  ~. \non \\
	\label{em_reg_dyn}
\end{eqnarray}
The square brackets in the first term on the right side denote the coincidence limit  $\tau^\prime \rightarrow \tau$ of the enclosed quantity along the unique geodesic connecting $\bar{z}^\alpha$ and $\bar{z}^{\alpha^\prime}$. Using the relation 
\begin{eqnarray}
	[ (\tau - \tau^\prime )_{; \beta} ] = - \bar{u}_\beta
\end{eqnarray}
which can be shown to be true from (\ref{sigma_vector}) in Appendix \ref{app_ql_exp}, this first term becomes
\begin{eqnarray}
	e^2 w_\mu ^{~\alpha \beta}  u_\beta \left( u_\alpha + [ V_{\alpha \gamma^\prime} ] u^\gamma \right) = \frac{e^2}{2} \, w_\mu ^{~\alpha} R_{\alpha \beta} u^\beta
\end{eqnarray}
where $w_\mu ^{~\alpha} = g_\mu ^{~\alpha} + u_\mu u^\alpha$ projects onto a direction orthogonal to the velocity $u^\mu$. Use has also been made of the coincidence limit of $V_{\alpha \gamma^\prime}$ \cite{Poisson}
\begin{eqnarray}
	\left[ V_{\alpha \gamma^\prime} \right] = - \frac{1}{2} \left( R_{\alpha \gamma} - \frac{1}{6} \, g_{\alpha \gamma} R \right)  ~.
\end{eqnarray}
 The last term on the right side of (\ref{em_reg_dyn}) is the tail term of the self-force with
\begin{eqnarray}
	A_{\alpha \beta} ^{tail} (z) = e \int _{\tau_i} ^\tau d\tau^\prime \, \nabla_\beta V_{\alpha \gamma^\prime} (z^\lambda, z^{\lambda^\prime} ) u^{\gamma^\prime} 
\end{eqnarray}
and is a history-dependent (non-Markovian) and non-local quantity that allows for radiation to backscatter off the background spacetime curvature and interact with the particle at a different place and time from the original emission. Because the Green's functions in a general spacetime are not known the contribution of the tail term to the self-force is quite difficult to compute. We will keep the tail term at this formal level in this paper. 

We now consider the diverging behavior of the local contribution to the self-force stemming from the direct part of the retarded Green's function.
The replacement (\ref{reg}) is a good approximation so long as $\Lambda^2 \sigma \gg 1$ as $\sigma \rightarrow 0$. This last condition allows the direct part of the retarded Green's function to be expanded near coincidence $\tau^\prime \rightarrow \tau$ in a quasilocal expansion so that the divergent behavior can be identified and dealt with. Appendix \ref{app_ql_exp} lists the quasilocal expansions of the relevant quantities appearing in the proper time integrand of (\ref{em_reg_dyn}). Using those results the mean particle dynamics becomes
\begin{eqnarray}
	 \left( m_o - e^2 g_{(1)}(r) \right) \bar{a}_\mu = F_\mu ^{ext} + f_\mu [ \bar{z}] 
	 \label{em_reg_cl_ALD}
\end{eqnarray}
where $F_\mu ^{ext}$ is some external force that may be present and $f_\mu$ is the self-force on the charge
\begin{eqnarray}	
	 f_\mu [\bar{z}] &=&  \frac{2 e^2}{3} \, g_{(2)}(r) w_\mu ^{~\alpha} [\bar{z}] \frac{ D \bar{a}_\alpha }{d\tau} \non \\
	 && + \frac{ e^2 }{ 6 } \! \left( 3 - c_{(1)}(r) \right) \! w_\mu ^{~\alpha } [\bar{z}] R_{\alpha \beta} (\bar{z}) \bar{u}^\beta \nonumber \\
	&&  + e \, w_\mu ^{~\alpha \beta} [\bar{z}] A_{\alpha \beta} ^{tail} (\bar{z}) + O(\Lambda^{-1})
	\label{em_reg_sf}
\end{eqnarray}
where the functions $c_{(n)}$ and $g_{(n)}$ are defined in Appendix \ref{coefficients}. Fig. \ref{fig_coeffs} shows their dependence on the elapsed proper time $r = \tau- \tau_i$ . These coefficients describe the time-dependence of the system (particle) degrees of freedom as the field modes with $\omega \lsim \Lambda$ become correlated with the particle after the initial time (when they were assumed to be uncorrelated).

\begin{figure} 
\includegraphics[height=6cm]{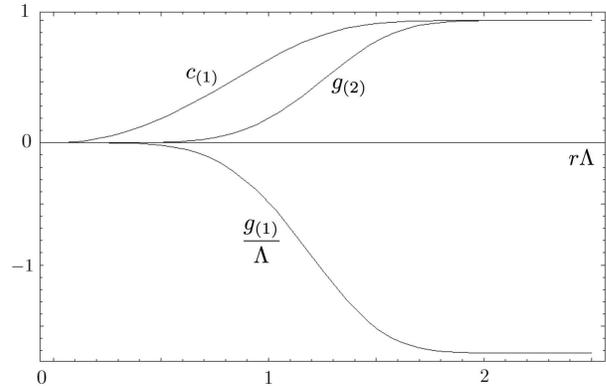}
    	\caption{Time-dependence of the coefficients appearing in (\ref{em_reg_cl_ALD}) and (\ref{em_reg_sf}). The function  $g_{(1)}$ has been divided through by $\Lambda$ so that it can be displayed on the same plot with $c_{(1)}$ and $g_{(2)}$.}
    \label{fig_coeffs}
\end{figure}

The terms proportional to $\Lambda^{-1}$ are irrelevant in the limit of infinitely large $\Lambda$ while those proportional to $\Lambda$ are relevant but divergent. From Appendix \ref{coefficients} it follows that the only divergent term above appears as a shift in the mass $m_o$. This shift $\delta m = -e^2 g_{(1)}(r)$ renormalizes the value of the (bare) mass $m_o$ to its physical value $m_{ph} = m_o + \delta m$. We therefore define $m_{ph}$ to be the physical mass of the point particle at rest. Notice that when $\Lambda \rightarrow \infty$ that the physical mass becomes independent of time and for a finite but large value of $\Lambda$ the shift in mass $\delta m$ tends to a numerical constant at late times. 

In the limit that $\Lambda$ goes to infinity $c_{(1)}$, $g_{(2)} \rightarrow 1$ and the classical particle equations become
\begin{eqnarray}
 	m_{ph} \bar{a}_\mu &=& \frac{2 e^2}{3} \, w_\mu ^{~\alpha} [\bar{z}] \frac{ D \bar{a}_\alpha }{d\tau} + \frac{e^2}{3} \, w_\mu ^{~\alpha } [\bar{z}] R_{\alpha \beta} (\bar{z}) \bar{u}^\beta \nonumber \\
 	&&  + e \, w_\mu ^{~\alpha \beta} [\bar{z}] A_{\alpha \beta} ^{tail} (\bar{z}) + F_\mu ^{ext} .
	\label{em_cl_ALD}
\end{eqnarray}
This is a generalization to curved spacetime of the well known Abraham-Lorentz-Dirac (ALD) equation describing the self-force of the electromagnetic field on the motion of a point electric charge \cite{DeWitt_Brehme}. 


The effective theory paradigm is a natural one to use when regularizing point particle divergences since one cannot expect classical electrodynamics (or even QED) to provide a valid description of the particle at arbitrarily small distances. So long as the typical distance scales being probed by an experiment are much larger than this size $1/\Lambda$ the particle behaves effectively like a point particle.

\subsection{Stochastic dynamics}
\label{em_stoch_dyn}

We now study the effects of the quantum field fluctuations (manifesting as classical stochastic forces) on the low-energy dynamics of the particle. We begin by computing the stochastic effective action
\begin{eqnarray}
	S_{SEA} [z,z^\prime] = \Re S_{CGEA} [z,z^\prime] - \xi^\mu \cdot j_\mu ^-
\end{eqnarray}
from (\ref{SEA}). Varying $S_{SEA}$ around the classical trajectory $\bar{z}^\mu$ to linear order in the world line fluctuations $\tilde{z}^\mu$ and performing a stationary phase approximation in the reduced density matrix gives the following Langevin equation
\begin{eqnarray}
	m_0 a_\mu = \vec{w}_\mu ^{~\alpha } [z] A_\alpha ^{ret} (z) + \eta_\mu [z]
\end{eqnarray}
where the stochastic force $\eta_\mu$ is related to the stochastic field $\xi (x)$ through
\begin{eqnarray}
	\eta_\mu [z] = e \, \vec{w}_{\mu}^{~\alpha} [z] \xi_\alpha (z) = - e \, w_\mu ^{~\alpha \beta} [z] 
	  \nabla_\beta \xi_\alpha(z)  ~.
\end{eqnarray}
As before, the retarded field $A_\alpha ^{ret}$ diverges and must be regularized. The prescription used in the previous section can be carried over so that the regularized ALD-Langevin equation is
\begin{eqnarray}
	\left( m_0 - e^2 g_{(1)}(r) \right) a_\mu = F_\mu ^{ext} (\tau) + f_\mu [z] + \eta_\mu [z]
	\label{ALD-Langevin}
\end{eqnarray}
where the (regulated) self-force is
\begin{eqnarray}
	f_\mu [z] &=&  \frac{2 e^2}{3} \, g_{(2)}(r) w_\mu ^{~\alpha } [z] \frac{ D a_\alpha }{d\tau} \non\\
	&& + \frac{e^2}{6} \left( 3 -  c_{(1)}(r) \right) w_\mu ^{~\alpha } [z] R_{\alpha \beta} (z) u^\beta  \nonumber \\
	&& + e \, w_\mu^{~\alpha \beta} [z] A_{\alpha \beta} ^{tail} (z) + O(\Lambda^{-1})  ~.
\end{eqnarray}
One must remember that these equations are only valid up to linear order in the fluctuations $\tilde{z}$ about the mean world line $\bar{z}$. Expanding the self-force $f_\mu$  in orders of the fluctuations using
\begin{eqnarray}
	f_\mu [z] = f_\mu [\bar{z}] + \int d\tau^\prime \, \tilde{z} ^{\nu^\prime} \left[ \frac{\delta}{\delta z^{\nu^\prime} } \, f_\mu [z^\alpha (\tau) ] \right] _{z=\bar{z}} + O (\tilde{z}^2) \non \\
	\label{sf_flucs}
\end{eqnarray}
and computing the linearization of those terms in the ALD-Langevin equation (\ref{ALD-Langevin})  involving the covariant $\tau$ derivatives (e.g. $a_\mu$) gives the following equation for the dynamics of the fluctuations
\begin{eqnarray}
	m_{\mu\nu} [\bar{z}] \ddot{\tilde{z}} ^\nu + \gamma_{\mu\nu} [\bar{z}] \dot{\tilde{z}}^\nu + \kappa _{\mu\nu} [\bar{z}] \tilde{z}^\nu \! &=& \! r_{\mu\nu} [\bar{z}] \dddot{\tilde{z}}^\nu + \eta_\mu [ \bar{z}] \non \\
	&+& \! O (\Lambda^{-1}, \tilde{z}^2 )  ~.
	\label{em_stoch_ALD}
\end{eqnarray}
Here we use overdots to denote $d/d\tau$. We have also used the semi-classical ALD equation for the mean world line (\ref{em_reg_cl_ALD}) in this derivation. The tensor coefficients $m_{\mu\nu}$, $\gamma_{\mu\nu}$, $\kappa_{\mu\nu}$ and $r_{\mu\nu}$ are given in Appendix \ref{stoch_coeff}.

Notice that (\ref{em_stoch_ALD}) is a linear differential equation for $\tilde{z}$ with a third derivative term and contains time-dependent coefficients that depend on the mean trajectory, which has non-Markovian behavior. While the mean motion is non-Markovian the stochastic fluctuations are Markovian in the sense that given a mean trajectory $\bar{z}$ the fluctuations do not depend on their own past history. 
The effective mass $m_{\mu\nu}$ for the fluctuations is not generically diagonal implying that the inertia of the fluctuations is not necessarily the same in all directions. This feature is exhibited in the other three tensor coefficients, which suggests that the fluctuations in one direction are linked with the fluctuations in the other spacetime directions.

\subsection{Comparisions with scalar field}

The scalar field case examined in \cite{GH1} contains interesting features that are not exhibitied in the electromagnetic one. There the effective mass of the scalar charge is found to be
\begin{eqnarray}
	m_\phi (\tau) = m_\phi ^{ph} - e^2 \int_{\tau_i}^\tau d\tau^\prime V ( \bar{z}^\lambda, \bar{z}^{\lambda^\prime} )
\end{eqnarray}
where $m_\phi ^{ph}$ is the renormalized mass of the particle at late times. Notice that $m_\phi$ is dependent on the history of the particle's classical motion via the non-local contribution from the tail term $\sim \int V$ so that the rate of change of the mass is
\begin{eqnarray}
	\frac{d m_\phi }{ d\tau} = - e^2 \bar{u}^\alpha \!\! \int ^\tau _{\tau_i} d\tau^\prime \, V_{; \alpha} 
\end{eqnarray}
whereas the renormalized mass of the electric charge is constant at late times. This feature of scalar charges has been studied in detail on some cosmological background spacetimes in \cite{Burko_Harte_Poisson} where they find, in the examples they consider, that the scalar mass can evaporate completely or can lose some of its rest mass only to regain it during later cosmological evolution. The scalar mass changes because the self-force $\sim e \partial_\mu \phi (z)$ is not orthogonal to the particle's four-velocity. The component of the self-force in the direction of $u^\alpha$ is precisely the quantity entering the rate of change of $m_\phi$ above. 
Choosing a different interaction for the scalar field, say $\sim e \int d\tau u^\mu \partial_\mu \phi$, generates a self-force orthogonal to the 4-velocity whence the mass $m_\phi$ remains constant. 

This behavior of the time-dependent mass in the scalar case becomes manifest in the qualitative description of the equations describing the dynamics of the world line fluctuations. Whereas these fluctuations in the electromagnetic case (\ref{em_stoch_ALD}) are Markovian, since they do not depend on their own past history, the fluctuations in the scalar case evolve in a non-Markovian way because of the appearance of
\begin{eqnarray}
	-e^2 \!\!\! \int d\tau^\prime \! \left( \tilde{z}^{\alpha} \vec{w}_\mu [\bar{z}] V_{; \alpha} ( \bar{z}^\lambda, \bar{z}^{\lambda^\prime} ) + \tilde{z}^{\alpha^\prime} \vec{w}_\mu [\bar{z}] V_{; \alpha ^\prime} (\bar{z}^\lambda, \bar{z}^{\lambda^\prime} ) \right) \non
\end{eqnarray}
in the dynamics, where $\vec{w}_\mu = a_\mu + w_\mu ^{~\alpha} \nabla_\alpha$. This term arises from linearizing the time-dependent mass $m_\phi (\tau)$ around the classical trajectory $\bar{z}^\mu$. Notice that the integration over the fluctuations at past times is weighted against the gradient of the tail so that this behavior disappears in Minkowski spacetime. If the classical motion is approximately geodesic then this term would seem to contribute little to the fluctuation dynamics unless the integration provides a significant cumulative (i.e. secular) contribution at late times. This non-Markovian term does not appear when the charge current couples to the field's gradient $\sim e \int d\tau u^\mu \partial_\mu \phi$.

\section{Linearized Metric Perturbations} 
\label{grav}

In this section we will study the motion of a small point mass $m_0$ moving through a linearized quantum metric perturbation field with $\Phi_A (x) = h_{\mu\nu}(x)$ in a curved vacuum background spacetime. The field and interaction actions describing the particle-field dynamics are 
\begin{eqnarray}
	&& S_E [h_{\mu\nu}] = - \frac{1}{64\pi} \int d^4x \, \sqrt{-g} \, h^{\mu\nu} \left( 2 \nabla ^\alpha \nabla _{(\mu} h_{\nu) \alpha}  \right. \nonumber \\
	&& {\hskip0.5in} \left. - \Box h_{\mu\nu} - \nabla_\mu \nabla _\nu h - g_{\mu\nu} \left( \nabla ^\alpha \nabla ^\beta h_{\alpha \beta} - \Box h \right) \right) \nonumber \\
	&& S_{int} [z, h_{\mu\nu}] = \frac{1}{2} \int d^4 x \, \sqrt{-g} \, h_{\mu\nu} T^{\mu\nu} (x;z) 
\end{eqnarray}
where 
\begin{eqnarray}
	\,T^{\mu\nu} (x;z) = m_o \int d\lambda \, \frac{ \delta^4 (x-z) }{ \sqrt{-g} } \, \frac{g^\mu _{~\alpha} (x,z) g^\nu _{ ~\beta} (x,z) u^\alpha u^\beta}{\sqrt{-u^\gamma g_{\gamma\delta} u^\delta } } . \nonumber \\
\end{eqnarray}
is the particle's stress-energy tensor. It is assumed that the lowest order particle dynamics describes a geodesic on the background vacuum spacetime so that the acceleration is of the order of the (infinitesimally small) mass $m_o$. (This assumption will be used repeatedly throughout the remainder.) We choose the Lorentz gauge for the trace-reversed metric perturbation $\gamma_{\mu\nu} = h_{\mu\nu} - \frac{1}{2} g_{\mu\nu} h$ using the gauge-fixing action
\begin{eqnarray}
	S_{gf} [h_{\mu\nu}] = - \frac{1}{32 \pi} \int d^4x \, \sqrt{-g} \, G^\mu G_\mu
\end{eqnarray}
where $G_\mu = \nabla^\nu \gamma_{\mu\nu}$ is the gauge-fixing function. From these we calculate the influence functional and find
\begin{eqnarray}
	F[z,z^\prime] &=& e^{- (1/4 \hbar) T_{\alpha \beta} ^- \cdot ( 8 \pi G_H ^{\alpha \beta \gamma^\prime \delta^\prime} ) \cdot T_{\gamma^\prime \delta^\prime}^- } \non \\
	&& \times e^{ (i/\hbar ) T_{\alpha \beta}^- \cdot ( 8 \pi G_{ret} ^{\alpha \beta \gamma^\prime \delta^\prime} ) \cdot T_{\gamma^\prime \delta^\prime}^+ }  ~.
	\label{grav_inf}
\end{eqnarray}
(The factor of $8 \pi$ multiplying the Green's functions is a result of the numerical factors appearing in $S_E$ and $S_{int}$.)
The retarded linearized metric perturbation is $h^{ret}_{\alpha \beta} = 16 \pi G^{ret}_{\alpha \beta \gamma^\prime \delta^\prime} \cdot T^{\gamma^\prime \delta^\prime}$ where the retarded Green's function solves
\begin{eqnarray}
	&& \Box G_{\alpha \beta} ^{~~~\gamma^\prime \delta^\prime} - \frac{1}{2} \, g_{\alpha \beta} \Box G_\epsilon ^{~\epsilon \gamma^\prime \delta^\prime} + 2 R_{\alpha ~ \beta} ^{~\, \mu ~\, \nu} G_{\mu \nu} ^{~~~ \gamma^\prime \delta^\prime } \non  \\
	&& {\hskip0.2in} = -2 \, g_{(\alpha} ^{~~\gamma^\prime} g_{\beta )} ^{~~\delta^\prime} \frac{\delta^4 (x-x^\prime)}{ \sqrt{-g} }
\end{eqnarray}
(The factor of 2 on the right side is a result of the symmetrization $g_{(\alpha} ^{~~\gamma^\prime} g_{\beta )} ^{~~\delta^\prime}$). In terms of the quantum two-point functions the Hadamard and retarded Green's functions are
\begin{eqnarray}
	8 \pi G^H _{\alpha \beta \gamma^\prime \delta^\prime } (x,x^\prime) &=& \big\langle\{ \hat{ h}_{\alpha \beta} (x) , \hat{h}_{\gamma^\prime \delta^\prime} (x^\prime) \} \big\rangle \non \\
	8 \pi G^{ret} _{\alpha \beta \gamma^\prime \delta^\prime} (x,x^\prime) &=& i \theta_+ (x,\Sigma) \big\langle [ \hat{h}_{\alpha \beta} (x), \hat{h} _{\gamma^\prime \delta^\prime} (x^\prime) ] \big\rangle \non
\end{eqnarray}
in the interaction picture.

\subsection{Semiclassical dynamics}

Using (\ref{cgea}) and (\ref{grav_inf}) we calculate the coarse-grained effective action
\begin{eqnarray}
	&& S_{CGEA} [z,z^\prime] = S_S [z] - S_S [z^\prime] \non \\
	&& {\hskip0.1in} + \frac{1}{2} \, T_{\alpha \beta}^- \cdot G_{ret} ^{\alpha \beta \gamma^\prime \delta^\prime} \cdot T_{\gamma^\prime \delta^\prime} ^+ + \frac{i}{2} \, T_{\alpha \beta} ^- \cdot G_H ^{\alpha \beta \gamma^\prime \delta^\prime} \cdot T_{\gamma^\prime \delta^\prime } ^- . \non \\
	\label{em_CGEA}
\end{eqnarray}
If the quantum field fluctuations provide a strong enough mechanism for decoherence then varying the CGEA with respect to $z^\mu _- = z^\mu - z^{\mu^\prime}$ gives the semiclassical equations of motion for the particle world line, which are
\begin{eqnarray}
	\bar{a}_\mu = \vec{w}_\mu ^{~\alpha \beta} [\bar{z}] \, h_{\alpha \beta}^{ret} (\bar{z})
	\label{gr_cl_eom}
\end{eqnarray}
where the derivative operator $\vec{w}_\mu ^{~\alpha \beta}$ here is
\begin{eqnarray}
	\vec{w}_\mu ^{~\alpha \beta} = w_\mu ^{~\alpha \beta \gamma} \nabla_\gamma = \left( \frac{1}{2} \, u^\alpha u^\beta w_\mu ^{~\gamma} - w_\mu ^{~(\alpha} u^{\beta)} u^\gamma \right) \nabla_\gamma .\non \\
\end{eqnarray}
It is computed from the variation of half the stress tensor with respect to the world line coordinates
\begin{eqnarray}
	\frac{1}{2} \frac{ \delta }{ \delta z^\mu (\tau) } \int d^4x \, \sqrt{-g} \, T^{\alpha \beta} (x; z) f(x) = \vec{w}_\mu ^{~\alpha \beta} [z] f(z) \non
\end{eqnarray}
where $f(x)$ is some suitable test function. As before, the quantity $w_\mu ^{~\alpha} = g_\mu ^{~\alpha } + u_\mu u^\alpha$ projects vectors in the direction of the 4-acceleration, orthogonal to $u^\alpha$.

It is more convenient to use the retarded Green's function associated with the trace-reversed perturbation $\gamma_{\mu\nu}$ rather than the metric perturbation $h_{\mu\nu}$ itself since Hadamard's ansatz can be applied directly to the former. In particular, if $\check{G}^{ret} _{\alpha \beta \gamma^\prime \delta^\prime}$ is the Green's function for the trace-reversed perturbation then Hadamard's ansatz (see \cite{Poisson}) gives
\begin{eqnarray}
	\check{G}^{ret \, \gamma^\prime \delta^\prime} _{\alpha \beta} (x,x^\prime) &=& 2 g_{( \alpha } ^{~~\gamma^\prime }  g_{\beta) } ^{ ~~ \delta^\prime}  \Delta^{1/2} (x,x^\prime) \delta_+ ( \sigma(x,x^\prime) ) \nonumber \\
	&&  + V_{\alpha \beta } ^{ ~~\gamma^\prime \delta^\prime} (x,x^\prime) \theta_+ (-\sigma (x,x^\prime) )
\end{eqnarray}
and $V_{\alpha \beta \gamma^\prime \delta^\prime}$ satisfies the homogeneous field equation
\begin{eqnarray}		\Box V_{\alpha \beta \gamma^\prime \delta^\prime } + 2 R_{\alpha ~ \beta} ^{~\, \mu ~\, \nu } V_{\mu \nu \gamma^\prime \delta^\prime} = 0
\end{eqnarray}
with characteristic data provided by its values on the forward light cone from $x^\prime$ \cite{Poisson}. 

In terms of $\gamma_{\mu\nu}$, (\ref{gr_cl_eom}) becomes
\begin{eqnarray}
	\bar{a}_\mu = \vec{w}_\mu ^{~\alpha \beta} [\bar{z} ] \, \left( \gamma ^{ret} _{\alpha \beta} (\bar{z}) - \frac{1}{2} \, g_{\alpha \beta} \gamma ^{ret} (\bar{z}) \right)
	\label{trrev_cl_eom}
\end{eqnarray}
where $\gamma ^{ret}= g^{\mu\nu} \gamma_{\mu\nu} ^{ret}$. As with the electromagnetic Green's function, the direct part of $\check{G}^{ret} _{\alpha \beta \gamma^\prime \delta^\prime}$ gives rise to a divergence when the two points are light-like separated. The self-force in (\ref{trrev_cl_eom}) requires the retarded Green's function to be evaluated along the particle's trajectory, which is time-like, so that the only contribution to the self-force occurs at coincidence when the two points are equal. In the next section, this divergence is regulated using our approach inspired by effective field theory.

\subsection{Regularization}

Introduce a regulator $\Lambda$ through the replacement (\ref{reg}) such that for particle energies much lower than this scale an effective description of the particle dynamics can be given without recourse to information about the high energy physics that is being ignored. We will begin by examining the first term on the right side of (\ref{trrev_cl_eom}).  After passing the derivative through the integral and remembering the assumption that $\bar{a}_\mu = O(m_o)$ it follows that
\begin{eqnarray}
	&& {\hskip-0.25in} \vec{w}_\mu ^{~\alpha \beta} \gamma_{\alpha \beta} ^{ret} (\bar{z}) \non \\
	&=& m_o w_\mu ^{~\alpha \beta \gamma} \left[ (\tau-\tau_i )_{;\gamma} \right] \left[ \bar{u}_\alpha \bar{u}_\beta \Delta^{1/2} \delta_\Lambda + V_{\alpha \beta \gamma^\prime \delta^\prime } \bar{u}^{\gamma^\prime } \bar{u}^{\delta^\prime} \right] \non \\
	&& + \frac{ m_o }{ 2 } \, w_\mu ^{~\gamma}  \int _{\tau_i} ^\tau d\tau^\prime \left\{ \delta_\Lambda (\sigma) \nabla_\gamma \Delta^{1/2} + \Delta^{1/2} \nabla_\gamma \delta_\Lambda (\sigma) \right\} \non \\
	&& + w_\mu ^{~\alpha \beta\gamma} \gamma_{\alpha \beta \gamma} ^{tail} (\bar{z} ) + O(m_o^2) ~.\non \\ 
	\label{reg_gamma}
\end{eqnarray}
The first term in the square brackets on the right side vanishes since $[ (\tau - \tau_i)_{; \gamma} ] = - \bar{u}_\gamma$ and $w_\mu ^{~\alpha \beta \gamma} \bar{u}_\alpha \bar{u}_\beta \bar{u} _\gamma = 0$. The coincidence limit of $V_{\alpha \beta \gamma^\prime \delta^\prime}$ is proportional to $R_{\gamma^\prime (\alpha  \beta ) \delta ^\prime}$ and vanishes when contracted against three velocity factors. The last term on the right side is the tail term of the self force with
\begin{eqnarray}
	\gamma_{\alpha \beta \gamma} ^{tail} (z) = m_o \int _{\tau_i} ^\tau d\tau^\prime \, \nabla_\gamma V_{\alpha \beta \gamma^\prime \delta^\prime}  u^{\gamma^\prime} u^{\delta^\prime}
\end{eqnarray}
describing the history-dependent effects of radiation emitted in the past that gets back-scattered onto the particle by the background curvature. It is not our intent here to calculate this contribution to the self-force. Instead, we focus on regularizing the divergence stemming from the direct part of the retarded Green's function.

Recall that the replacement $\delta(\sigma) \rightarrow \delta_\Lambda(\sigma)$ of (\ref{reg}) is a good approximation if $\Lambda^2 \sigma \gg 1$ as $\sigma$ becomes vanishingly small, which allows the direct part of the Green's function to be expanded near coincidence as $\tau^\prime \rightarrow \tau$ in order to identify the divergence. Using the quasi-local expansions of the van-Vleck determinant and the other quantities appearing on the right side of (\ref{reg_gamma}), listed in Appendix \ref{app_ql_exp},  and recognizing that $\bar{a}_\mu = O(m_o)$, it follows that
\begin{eqnarray}
	{\hskip-0.25in} \vec{w}_\mu ^{~\alpha \beta} \gamma_{\alpha \beta} ^{ret} (\bar{z}) &=& - m_o w_\mu ^{~\alpha} [\bar{z}] R_{\alpha \beta} (\bar{z}) \bar{u}^\beta c_{(1)}(r) \non \\
	&& + w_\mu ^{~\alpha \beta \gamma} [\bar{z}] \gamma_{\alpha \beta \gamma} ^{tail} (\bar{z})+ O( \Lambda^{-1} ) .
\end{eqnarray}
The background spacetime is vacuous so that the Ricci tensor vanishes and the only contribution  comes from the tail term
\begin{eqnarray}
	\vec{w}_\mu ^{~\alpha \beta} \gamma_{\alpha \beta} ^{ret} (\bar{z}) = w_\mu ^{~\alpha \beta \gamma} [\bar{z}] \gamma_{\alpha \beta \gamma} ^{tail} (\bar{z})+ O(\Lambda^{-1} ) .
\end{eqnarray}
Performing a similar computation for $\vec{w}_\mu ^{~\alpha \beta} (g_{\alpha \beta} \gamma^{ret} )$ results in
\begin{eqnarray}
	\vec{w}_\mu ^{~\alpha \beta} (g_{\alpha \beta} \gamma^{ret} ) = w_\mu ^{~\alpha \beta \gamma} [\bar{z}] g_{\alpha \beta} (\bar{z}) \gamma_{\delta \epsilon \gamma} ^{tail} (\bar{z}) g^{\delta \epsilon} + O(\Lambda^{-1}) . \non \\
\end{eqnarray}

Putting into (\ref{trrev_cl_eom}) these last two equations together give the MSTQW equation describing the self-force on the point mass 
\begin{eqnarray}
	m_o \bar{a}_\mu = m_o w_\mu ^{~\alpha \beta \gamma} [\bar{z}] \, h_{\alpha \beta \gamma} ^{tail} (\bar{z}) + O( \Lambda^{-1})
	\label{grav_reg_cl_eom}
\end{eqnarray}
where 
\begin{eqnarray}
	h_{\alpha \beta \gamma} ^{tail} = \gamma_{\alpha \beta \gamma} ^{tail} - \frac{1}{2} \, g_{\alpha \beta} g^{\delta \epsilon} \gamma ^{tail} _{\delta \epsilon \gamma} .
	\label{tail}
\end{eqnarray}
Notice that the self-force of the particle is determined solely by the tail term of the linearized perturbation $h_{\mu\nu}$. This makes solving for the particle's motion very difficult as its dynamics are highly nonlinear and history-dependent, requiring an integration over past times and positions.


\subsection{Stochastic dynamics}
\label{grav_stoch_dyn}

We now turn to study the effects of the quantum field fluctuations of the linearized metric perturbations on the low-energy particle dynamics. In particular, we are interested in the attributes and consequences of the quantum field fluctuations manifesting as classical stochastic forces on the particle. From the influence functional (\ref{grav_inf}) and the discussion in Section \ref{stoch_formalism} one can show that the stochastic effective action for this system and environment is
\begin{eqnarray}
	S_{SEA} [z,z^\prime] = \Re S_{CGEA} [z,z^\prime] - \xi^{\mu\nu} \cdot j_{\mu\nu}^-
\end{eqnarray}
where the current density is $j_{\mu\nu}^- = T_{\mu\nu}^- /2$ and the CGEA is given by (\ref{em_CGEA}). Assuming sufficiently strong decoherence, expanding the SEA about the mean trajectory and performing a stationary phase approximation in the resulting reduced density matrix gives rise to the stochastic particle dynamics
\begin{eqnarray}
	m_o a_\mu = m_o \vec{w}_\mu ^{~\alpha \beta} [z] \, h_{\alpha \beta} ^{ret} (z) + \eta_\mu [z]
\end{eqnarray}
where the stochastic force $\eta_\mu$ is related to the stochastic field $\xi_{\mu\nu}(x)$ through
\begin{eqnarray}
	\!\!\! \eta_\mu = m_o \vec{w}_\mu^{~\alpha \beta} [z] \, \xi _{\alpha \beta} (z) = - m_o w_\mu ^{~\alpha \beta \gamma}  [z] \nabla _\gamma \xi_{\alpha \beta} (z) .
\end{eqnarray}
As before, the retarded field $h_{\alpha \beta}^{ret}$ diverges and must be regularized. The prescription used in the previous section can be applied here so that the regularized MSTQW-Langevin equation reads
\begin{eqnarray}
	m_o a_\mu = f_\mu [z] + \eta_\mu[z]
	\label{MSTQW-Langevin}
\end{eqnarray}
where the (regulated) self-force is
\begin{eqnarray}
	f_\mu [z] = m_o w_\mu^{~\alpha \beta \gamma}[z] \, h_{\alpha \beta \gamma} ^{tail} (z) +O(\Lambda^{-1})
\end{eqnarray}
and the tail term is given in terms of the trace-reversed perturbations in (\ref{tail}). We do not include an external force here as we did for the electromagnetic case since the lowest order motion of astrophysical bodies is geodesic on the background spacetime. Also, these equations are only valid up to linear order in the fluctuations $\tilde{z}$ about the mean world line $\bar{z}$.

Expanding the self-force $f_\mu$ in orders of the fluctuations using (\ref{sf_flucs}) and computing the linearization of those terms involving the covariant $\tau$ derivatives (e.g. $a_\mu$) gives the equation for the dynamics of the fluctuations
\begin{eqnarray}
	m_{\mu\nu} [\bar{z}] \ddot{\tilde{z}}^\nu + \gamma_{\mu\nu} [\bar{z}] \dot{\tilde{z}}^\nu + \kappa_{\mu\nu} [\bar{z}] \tilde{z}^\nu = \eta_\mu [\bar{z}] + O(\Lambda^{-1}, \tilde{z}^2) . \non \\
	\label{grav_stoch_ALD}
\end{eqnarray}
We have also used the semi-classical dynamics for the mean world line (\ref{grav_reg_cl_eom}) in this computation. The tensor coefficients $m_{\mu\nu}$, $\gamma_{\mu\nu}$ and $\kappa_{\mu\nu}$ are given in Appendix \ref{stoch_coeff}. In particular we note that $m_{\mu\nu} \propto g_{\mu\nu}$. In flat spacetime one can show that (\ref{grav_stoch_ALD}) becomes in Cartesian coordinates
\begin{eqnarray}
	m_o \ddot{\tilde{z}}_\mu = \eta_\mu [\bar{z}] + O(\Lambda^{-1}, \tilde{z}^2)
\end{eqnarray}
since the tail term vanishes identically in Minkowski spacetime. At $O(\Lambda^0)$ there is no dissipation term appearing here, which implies that the two-point function of $\dot{\tilde{z}}$ and $\tilde{z}$ could grow unbounded in time in the strict point-particle limit $\Lambda \rightarrow \infty$. However, if $\Lambda$ is large but finite then a simple scaling argument implies that dissipation effects from the neglected $O(\Lambda^{-1})$ terms could begin to appear on a time scale $\sim \Lambda$. Dissipation from higher order terms arising from the nonlinearities of the full metric perturbation field equations might begin to appear on a time scale $\sim 1/m_0$.

As with the electromagnetic case earlier, (\ref{grav_stoch_ALD}) is a linear differential equation for $\tilde{z}$. The important difference is that the third derivative of $\tilde{z}$ gives no contribution at this order. Hence, only the initial position and velocity of the fluctuations are sufficient to obtain a unique solution. This is unlike the electromagnetic case discussed earlier since one needs to introduce an external force to obtain unambiguous, runaway-free solutions. 

The world line fluctuations may be computed from a complete knowledge of the mean trajectory $\bar{z}$. While the semi-classical motion of $\bar{z}$ is non-Markovian generically, depending on its past motion to determine its present state, the fluctuations are Markovian in the sense that there is no integration over the past behavior of $\tilde{z}$.

\section{Phenomenological noise and noise-induced drift}

The noise $\eta_\mu [\bar{z}]$ in the Langevin equations describing the stochastic motion of the particle in (\ref{em_stoch_ALD}) and (\ref{grav_stoch_ALD}) are obtained from coarse-graining the environment. 
Our derivations of the equations of motion for the mean and fluctuation world lines assume a closed system to begin with. This treatment has the distinct advantage that it can preserve the self-consistency between the system and the environment in considering the effects of back-reaction. However, in many circumstances the stochastic dynamics of a system is treated phenomenologically with a noise put in by hand. Quite generally, for a tensor field $\Phi_A(x)$ this description follows from the stochastic equation
\begin{eqnarray}
	m a_\mu = F_\mu ^{ext}  + f_\mu[z] + \eta_\mu ^+
	\label{phenom_stoch_ALD}
\end{eqnarray}
where $F_\mu ^{ext}$ is some external force and $f_\mu$ is the self-force on the particle arising from its (non-local and history-dependent) interaction with the retarded field $\Phi_A ^{ret}$. We denote the phenomenological noise $\eta_\mu^+$ put in by hand by a $+$ superscript to distinguish them from those {\it derived} from first-principles considerations. This add-on stochastic force could have a classical origin (e.g. high temperature thermal fluctuations of a bath) or it could have no known single identifiable origin. Furthermore, since the $\eta_\mu ^+$ is not derived from an initially closed system it is likely to be inconsistent with the dynamics of the trajectory by  failing to satisfy a fluctuation-dissipation relation for world line displacements around an equilibrium trajectory (see, e.g., \cite{HPZ}). In a phenomenological treatment one also needs to {\it specify} the noise kernel $\langle \eta_\mu ^+ (\tau) \eta_{\nu^\prime} ^+ (\tau^\prime) \rangle_{\eta^+}$ befitting the model, rather than {\it deriving} it.


Keeping this defect in mind (i.e., no guarantee of consistency) the analysis of Sections \ref{em_stoch_dyn} and \ref{grav_stoch_dyn} carry over for a phenomenological noise. Given any kind of noise the equations of motion for the fluctuations around the mean trajectory of the particle moving through a (classical) field subjected to the self-force from radiation reaction is given by
\begin{eqnarray}
	m_{\mu\nu} [\bar{z}] \ddot{\tilde{z}} ^\nu + \gamma_{\mu\nu} [\bar{z}] \dot{\tilde{z}}^\nu + \kappa _{\mu\nu} [\bar{z}] \tilde{z}^\nu &=& r_{\mu\nu} [\bar{z}] \dddot{\tilde{z}}^\nu + \eta_\mu ^+ \non \\
	&& + O (\Lambda^{-1}, \tilde{z}^2 )
\end{eqnarray}
where the tensor coefficients for the respective electromagnetic and gravitational (with $r_{\mu\nu}=O(\Lambda^{-1})$) cases are given in Appendix \ref{stoch_coeff}. 


It is interesting to observe that when some source of noise acts as a stochastic force on the particle that a noise-induced drifting motion results. In particular, the stochastic force on the particle causes it to undergo rapid motions that enables the particle to experience different values of the inhomogeneous external fields. Averaging over the stochastic fluctuations results in a noise-induced drift that depends on the correlations of the stochastic force and the gradients of the external fields. 
Clearly, the noise-induced drift is a second-order effect in terms of the world line fluctuations. In this discussion we do not need to worry about quantum corrections from higher-order loops in the effective action, as the noise is not necessarily of a quantum origin, so we can expand (\ref{ALD-Langevin}) and (\ref{MSTQW-Langevin}) beyond the linear order in the fluctuations of the particle trajectory. A similar behavior is encountered frequently in plasma physics where the time-averaged Larmor motion results in a net velocity drift if the charge is moving through an inhomogeneous external magnetic field \cite{PlasmaBook}.

In order to highlight the essential physics of the noise-induced drift we consider the non-relativistic motion of an electrically charged particle and a point mass moving through a flat background spacetime. Doing so allows us to focus on this particular issue rather than on more complex and subtle issues that arise in the fully relativistic problem or in curved spacetime. Furthermore, the drifts commonly encountered in plasma physics are described by charges moving non-relativistically.

\subsection{Electromagnetic noise-induced drift}

To find the noise-induced drift in the electromagnetic case, we begin with the non-relativistic limit of (\ref{phenom_stoch_ALD}) describing the motion of an electric point charge moving through a flat background spacetime and coupled to a phenomenological stochastic force $\eta^+ (z)$
\begin{eqnarray}
	m a_i = e \left( E_i (z) + \epsilon _{ijk} u_j B_k (z) \right) + f_i [z] + \eta_\mu ^+ (z)
	\label{NR_phenom_ALDL}
\end{eqnarray}
where the external force $\vec{F}^{ext}$ is taken to be the usual Lorentz force. (The Latin indices take on the values from $1$ to $3$ and $\epsilon_{ijk}$ is the Levi-Civita totally antisymmetric tensor with $\epsilon_{123}=1$.) The self-force $f_i = 2 e^2/3c^3 \, \dot{a}_i$ contains no tail term here since there is no background curvature to back-scatter the emitted radiation. Substituting $z= z_0 + \delta z$ in (\ref{NR_phenom_ALDL}) and expanding in powers of the fluctuations gives
\begin{eqnarray}
	\!\!\! && m \left( a_{0i} + \delta a_i \right) = e E_i (z_0) + e \delta z^a \partial_a E_i (z_0) \non \\
	&& {\hskip0.1in} + \frac{e}{2} \delta z^a \delta z^b \partial_a \partial_b E_i (z_0) + e \epsilon _{ijk} u_{0j} B_k (z_0) \non \\
	&& {\hskip0.1in} + e \epsilon_{ijk} \delta u_j B_k (z_0) + e \epsilon_{ijk} u_{0j} \delta z^a \partial_a B_k (z_0) \non \\
	&& {\hskip0.1in} + e \epsilon_{ijk} \delta u_j \delta z^a \partial _a B_k (z_0) + \frac{e}{2} \epsilon_{ijk} u_{0j} \delta z^a \delta z^b \partial_a \partial_b B_k (z_0) \non \\
	&& {\hskip0.1in} + \frac{2e^2}{3c^3} \dot{a}_{0i} + \frac{2e^2}{3c^3} \dot{\delta a}_i + \eta_i ^+ (z_0) + \delta z^a \partial_a \eta_i ^+ (z_0) + \cdots \non \\
	\label{expanded_stoch_ALD}
\end{eqnarray}
We assume that the variations of the fields occur over distances much larger than $|\vec{\delta z}|$. The world line fluctuations are assumed very fast compared to the averaged motion so we expect that $|\vec{\delta u}| \gg | \vec{u} |$ and similarly for the accelerations. This allows us to make the approximation 
\begin{eqnarray}
	m \delta a_i - e \epsilon_{ijk} \delta u^j B^k (z_0) \approx \eta_i ^+ (z_0)
	\label{em_fluctuations}
\end{eqnarray}
where the stochastic force $\eta_i ^+$ drives the world line fluctuations. We assume that the typical time scale of the fluctuations $\Delta t$ is much larger than the time for light to cross the ``classical" size of the particle $\sim 2e^2/3mc^2$ so that the radiation reaction term, which is proportional to $\dot{\delta a}$, can be neglected. The equation of motion for $z_0$ then becomes, after taking the stochastic expectation value of (\ref{expanded_stoch_ALD}),
\begin{eqnarray}
	&& \!\! m a_{0i} \approx e \left( E_i (z_0) + \epsilon_{ijk} u_{0}^j B^k (z_0) \right) + \frac{2e^2}{3 c^3} \dot{a}_{0i} \non \\
	&& {\hskip0.1in} + \frac{e}{2} \big\langle \delta z^a \delta z^b \big\rangle \, \partial_a \partial_b E_i (z_0) + e \epsilon _{ijk} \big\langle \delta u^j \delta z^a \big\rangle \, \partial_a B^k (z_0) \non \\
	&& {\hskip0.1in} + \frac{e}{2} \epsilon _{ijk} u_0 ^j \big\langle \delta z^a \delta z^b \big\rangle \, \partial_a \partial_b B^k (z_0) + \big\langle \delta z^a \partial_a \eta_i ^+ (z_0) \big\rangle  \non \\
\end{eqnarray}
The terms involving the stochastic averages are defined as the noise-induced drift force so that
\begin{eqnarray}
	&& \!\!\!\!\!\!\! F_i ^{drift} \! = \frac{e}{2} \big\langle \delta z^a \delta z^b \big\rangle \, \partial_a \partial_b E_i (z_0) + e \epsilon _{ijk} \big\langle \delta u^j \delta z^a \big\rangle \, \partial_a B^k (z_0) \non \\
	&& {\hskip0.1in} + \frac{e}{2} \epsilon _{ijk} u_0 ^j \big\langle \delta z^a \delta z^b \big\rangle \, \partial_a \partial_b B^k (z_0) + \big\langle \delta z^a \partial_a \eta_i ^+ (z_0) \big\rangle \non \\
	\label{drift}
\end{eqnarray}
and
\begin{eqnarray}
	\!\!\! m \vec{a}_0 = e \vec{E} (z_0) + e \, \vec{u}_0 \! \times \! \vec{B} (z_0) + \frac{2e^2}{3c^3} \dot{\vec{a}}_0 + \vec{F} ^{drift} [z_0]
	\label{guiding_center}
\end{eqnarray}
The first three terms of the drift are a result of the variation of the external fields with position and the curvature of the external electromagnetic field lines. The last term of the drift results from the worldline fluctuation away from $z_0$ coupling to the variation of the stochastic force. If the stochastic force is independent of position then this term will vanish identically and any contribution to the noise-induced drift will result from variations in the applied electric and magnetic fields. 


To solve (\ref{guiding_center}) one needs the solution to (\ref{em_fluctuations}) for the fluctuations in terms of the stochastic force $\vec{\eta} ^+$. The solution is
\begin{eqnarray}
	\delta u_i = \frac{1}{m} \, (K^{-1}) _i ^{~a} \int _{t_i} ^t dt^\prime \, K_{a b} \eta ^b (z_0) 
\end{eqnarray}
where the homogeneous solution is ignored since we are interested in the effect of the noise on the averaged particle motion. The integrating factor $K_{ab}$ is
\begin{eqnarray}
	K_{ab} &=& \exp \left( - \frac{e}{m} \epsilon_{abm} \int _{t_i} ^t dt^\prime \, B^m (z_0 ^{\ell^\prime}) \right) \non \\
	&=& \delta_{ab} - \frac{e}{m} \epsilon_{abm} \int _{t_i} ^t dt^\prime \, B^{m} (z_0 ^{\ell^\prime} ) \non \\
	&& + \frac{ e^2 }{ 2 m^2} \epsilon_{arm} \epsilon^r _{~ bn} \int _{t_i } ^t dt^\prime \int _{t_i } ^{t^\prime} dt^{\prime\prime} \, B^{m} (z_0 ^{\ell^\prime} ) B^n (z_0 ^{\ell^{\prime\prime} } ) \non \\
	&& + \cdots
\end{eqnarray}
Integrating $\delta u$ over time gives the world line fluctuations $\delta z$. The terms in $\delta z$ not involving the stochastic force are ignored so that one is left with
\begin{eqnarray}
	 F_i ^{drift} \!\! &=& \!\! \int _{t_i} ^t \!\! dt_1 \int _{t_i} ^{t_1} \!\! dt_2 \int _{t_i} ^t \!\! dt_3 \int _{t_i} ^{t_3} \!\! dt_4   \left( F_{imn} + F_{imnp} \partial^p \right) \non \\
	&& {\hskip0.35in} \times \big\langle \eta^{+m} (z_0 ^{\ell_2}) \eta^{+n} (z_0 ^{\ell_4} ) \big\rangle_{\eta^+} \non \\
	\label{em_drift}
\end{eqnarray}
where the distributions $F_{imn}$ and $F_{imnp}$ are given by
\begin{eqnarray}
	&& \!\!\!\!\!\! F_{imn} = \frac{e}{2m^2} \partial_a \partial_b \left( E_i (z_0) + \epsilon_{ijk} u_0 ^j B^k (z_0) \right) \non \\
	&& {\hskip0.5in} \times (K^{-1} _1)^{br}  K_{2 r m} (K^{-1} _3 ) ^{as} K_{4 sn} \non \\
	&& \!\! + \frac{2 e}{m^2} \epsilon_{ijk} \partial_a B^k (z_0) \delta (t-t_1) (K^{-1} _1) ^{j r} K_{2 rm} (K^{-1} _3) ^{as} K_{4 sn} \non \\
	&& \!\!\!\!\!\! F_{imnp} = \frac{4}{m} \delta (t-t_1) \delta (t_1-t_2) \delta_{im} (K^{-1} _3 )^{~s} _p K_{4 sn}
\end{eqnarray}
where a subscript on the integrating factor $K$ refers to the designated time, e.g. $K_1 = K(t_1)$, etc. This expression for the drift force is then used to solve for the world line coordinates $\vec{z}_0$ in (\ref{guiding_center}). This is a difficult task given the nonlinear and non-Markovian behavior of the dynamics. The history-dependent contribution coming from the drift force requires a knowledge of $z_0$ and the stochastic correlation function for all times in the past. 

If the applied fields vary over a distance much larger than the Larmor radius then the usual drifts that occur in plasma physics can still be deduced from (\ref{guiding_center}). These drifts have been lumped into determining the motion of $z_0$ in order to isolate the new noise-induced drift $\vec{F} ^{drift}$ from the usual plasma physics drifts (e.g. grad-B drift, etc.).

A similar expression for the noise-induced drift determined above can be derived in our open quantum system framework. The quantum noise-induced drift would capture some of the information contained in the quantum fluctuations of the world line intrinsic to the particle and could provide information about the quantum statistical state of the environment and its influence on the system variables.
However, these considerations require a much more delicate analysis that goes beyond the scope of this paper and the stochastic semiclassical approximation adopted within. 

\subsection{Gravitational noise-induced drift}

A similar noise-induced drift seems to exist for the motion of a point mass moving through linearized metric perturbations in the presence of a source of noise. As before we consider the particle to be moving non-relativistically in Minkowski spacetime and the noise to be phenomenological and dependent on the particle's trajectory. Proceeding along a line similar to the electromagnetic case above we find that substituting $z = z_0 + \delta z$ into (\ref{phenom_stoch_ALD}) and expanding in orders of the world line fluctuations gives two equations. The first describes the particle fluctuations
\begin{eqnarray}
	m \delta a_i \approx \eta_i ^+ (z_0)
	\label{grav_fluctuations}
\end{eqnarray}
and the second describes the stochastically-averaged motion of the particle
\begin{eqnarray}
	m a_{0i} \approx  F_i ^{drift} [z_0]
\end{eqnarray}
where the noise-induced drift is
\begin{eqnarray}
	F_i ^{drift} \!\! &=& \big\langle \delta z^m \partial_m \eta_i ^+ (z_0) \big\rangle _{\eta^+} \non \\
	&& {\hskip-0.5in} = \frac{1}{m} \int_{t_i} ^t dt_1 \!\! \int _{t_i} ^{t_1} \!\! dt_2 \, \partial_m \big\langle \eta_i ^+ (z_0 ^{\ell}) \eta^{+m} (z_0 ^{\ell_2} ) \big\rangle_{\eta^+}
	\label{grav_drift}
\end{eqnarray}
The drift force then takes the same form as the electromagnetic drift with no applied fields acting on the particle. In particular, when the noise is independent of the particle's position there is no noise-induced drift.

\section{Validity of the Quasilocal expansion and order-reduction}

In the first half of this paper we used the influence functional formalism and various approximations to obtain the low-energy effective dynamics of the particle, both for its semiclassical (mean) and stochastic semiclassical (mean and fluctuations) motion. Here, the domain of validity of the quasilocal expansion and this semiclassical treatment will be discussed and compared with the relevant scales for weak and strong radiation damping.

In the effective field theory paradigm a regulator $\Lambda$ is introduced for controlling the ultraviolet divergences appearing in the direct part of the retarded fields such that $\Lambda ^2 \sigma \gg 1$ with $\sigma$ small and approaching zero. After expanding $\sigma$ near coincidence (see Appendix \ref{app_ql_exp}) the time scale of the quasi-local expansion $\Delta \tau = s$ is governed by $\Delta \tau \gg \Lambda^{-1}$. Recall that for elapsed times larger than $\sim \Lambda^{-1}$ the time-dependent coefficients $c_{(n)}$ and $g_{(n)}$ rapidly approach constant values. However, at the initial time these coefficients vanish giving the false impression that the usual problems of pre-acceleration and run-away solutions have been resolved. This time-dependence can be viewed as a consequence of choosing a factorized initial state in our derivations since it takes a time of order $\Lambda^{-1}$ for the field modes to fully interact with the particle as they were uncorrelated at the initial time. To better study the pre-acceleration and run-away solutions we should choose a more physical initial state than the factorizable one used in most treatments.

The mean and stochastic particle dynamics are obtained here by using the Gaussian approximation to compute the reduced density matrix, which amounts to working at the tree-level in both the particle and the field sectors. This implies that $\Delta \tau$ should be much longer than the time scale for creating particle pairs, $\Delta \tau \gg \hbar /m = \lambda_C$ where $\lambda_C$ is the particle's Compton wavelength. 

Another relevant scale appears when trying to find unique, physical solutions to the ALD equation (\ref{em_reg_cl_ALD}), which contains a term with a third derivative of the particle's position. As is well known, this term is responsible for the problematic existence of pre-accelerated, acausal and runaway solutions. These kinds of solutions can be eliminated if the self-force is weak compared to other forces acting on the particle. In particular, an asymptotic expansion in powers of $r_0 \sim 2e^2/3 mc^2$, called the Landau approximation or order-reduction \cite{LL}, is employed to obtain physical solutions that require only an initial position and velocity. The Landau approximation converts the ALD equation (of third order) to the so-called Landau-Lifshitz equation (of second order). See \cite{Rohrlich} for an interesting discussion of these equations and order-reduction. The quantity $r_0$ is often called the ``classical" size of the charge \cite{LL}. 

Using order-reduction, the lowest order solution is found by simply ignoring the self-force so that the radiation damping is assumed weak. The time-scale of the dynamics is then determined mostly by the external force so that if $F_\mu ^{ext}$ varies on a scale $\lambda_{ext}$ then $\Delta \tau \sim \lambda_{ext}$. In curved spacetime the self-force will be weak if $r_0 \ll \Delta \tau$ and the scale associated with the spacetime curvature $\lambda_R$ is much greater than $r_0$. 

For the electron, $\lambda_C =137 r_0 \gg r_0$ and one might choose to set $\Lambda^{-1} \sim \lambda_C$ to justify ignoring the effects of electron-positron pair production from appearing in the semi-classical particle dynamics. For an ion or a larger charged body, its physical size $R_0$ will undoubtedly dwarf its ``classical" size $r_0$ and Compton wavelength $\lambda_C$ so that one might choose $\Lambda^{-1} \sim R_0$ in order to ignore any effect resulting from the object's finite spatial extent and describe the object effectively as a point particle. Our approach would need to be augmented if we wished to include the effects of extended charged bodies. One way to do this while still maintaining the effective field theory paradigm is to include all possible terms into the point-particle action $S_S[z]$ that are consistent with reparametrization and diffeomorphism invariance. This provides a model-independent way to parametrize the contributions to the dynamics from the body's size. The couplings of these extra terms can then be determined by matching this effective theory to the theory describing the body on microscopic scales. See \cite{Goldberger_Rothstein} which takes a similar approach to construct a framework to derive the post-Newtonian equations describing the motions of neutral (spinning) extended bodies interacting gravitationally.

For the gravitational case, a third derivative of the particle's position appears at order $O(\Lambda^{-1}, m_0)$ so one does not need to implement the order-reduction procedure to find well-behaved solutions to (\ref{grav_reg_cl_eom}). One may be interested in studying the motion of a small mass black hole moving in the spacetime of a black hole of much greater mass, e.g. at least for obtaining gravitational waveforms for the LISA experiment. In this case, their gravitational radii are sufficiently different to justify describing the smaller body as a point particle. If the smaller body is a neutron star then a description that could take into account the star's finite extent, rapid spinning and strong magnetic fields might be more desirable. In such a scenario one might wish to take $\Lambda ^{-1} \sim R_{ns}$ where $R_{ns}$ is the size of the neutron star. 
In \cite{Goldberger_Rothstein} the authors used such an effective point particle theory to describe the motion of spinning, slowly moving and weakly gravitating extended bodies. Among other interesting results they found that in certain cases the extended bodies can be treated effectively and accurately as point particles up to $O(v^{10})$ in their relative velocity.

\section{Summary}

In this paper we have given a first-principles derivation of the electromagnetic ALD equation for the world line of an electrically charged point particle interacting with a  quantum vector field $A_\mu$ as it moves in a given background curved spacetime. Our Eq. (\ref{em_cl_ALD}) for the low-energy effective particle dynamics agrees with the results obtained earlier \cite{DeWitt_Brehme}. In a similar vein, we also derived from first principles the MSTQW equation \cite{MST, Quinn_Wald} describing the motion of a point mass moving through a vacuous background spacetime, including the history-dependent self-force experienced by the body. If the quantum fluctuations in the field strongly decohere quantum histories of the particle then the particle's quantum fluctuations can be ignored and the semiclassical motion (\ref{em_reg_cl_ALD}) and (\ref{grav_reg_cl_eom}) obtained. We use an effective field theory method to regulate the singular behavior of the retarded field. For a sufficiently fine temporal resolution but still much coarser than the inverse of our ultraviolet regulator $\Lambda$, a quasilocal expansion can be used to obtain the contributions relevant to the self-force and those that are irrelevant in the infinite $\Lambda$ limit. (For large and finite values of $\Lambda$ the $O(\Lambda^{-1})$ terms are dependent on the particular regularization scheme implemented as well as the details of the microscopic theory that it must be matched to.) This approach provides an intuitive and effective way to identify and regularize the divergences appearing in the self-force.



Fluctuations in the quantum field are expected to affect the
particle's motion causing it to fluctuate by an amount $\tilde{z}$
around the mean trajectory $\bar{z}$ given by the solutions to the
semi-classical equation. We derived the correlations of the
stochastic force $\eta_\mu[\bar{z}]$ associated with the
fluctuations of the quantum field, and a Langevin
equation for both the electromagnetic and gravitational cases. The dynamics of the fluctuation in the trajectory $\tilde{z}$ is Markovian but depends on the non-Markovian behavior of the mean trajectory $\bar{z}$. This is to be contrasted with the scalar field case (\cite{GH1}) where it is found that the fluctuations $\tilde{z}$ do depend on their own history in a curved spacetime for the particular particle-field coupling chosen.

Instead of the noise $\eta_\mu [\bar{z}]$ derived here, from first-principles as related to the fluctuations of
quantum fields, one can consider some other classical noise
$\eta_\mu ^+$ suitably chosen to model some stochastic source in a
phenomenological description. We can still use
(\ref{em_stoch_ALD}) and (\ref{grav_stoch_ALD}) with $\eta_\mu [\bar{z}]$ replaced by $\eta_\mu ^+$ to study the effect of such noises on the fluctuations of the
particle trajectory. Since the origin for
the noise is no longer due to a quantum field we need not worry
about keeping up to linear order in $\tilde{z}$ in these Langevin equations. Instead, in flat spacetime and for slowly moving particles, expanding the solution in terms of a slowly varying world line about which fast fluctuations occur $z = z_0 + \delta z$ and taking the stochastic average (expectation value) of the slow motion dynamics we show that there is, in general, a non-vanishing noise-induced drift force (\ref{em_drift}), (\ref{grav_drift}) resulting from the correlations of the stochastic force. Instead of using phenomenological noise, we can instead derive this noise-induced drift effect of the quantum field fluctuations using the influence functional. Such an effect, while undoubtedly small, contains information about the quantum statistics of the field that might have observational consequences for experiments involving the motion of particles in the presence of boundary-constrained quantum fields (e.g. Casimir and Casimir-Polder effects).

\begin{acknowledgements}
	We thank Phil Johnson and Albert Roura for useful discussions. This work is supported in part by NSF grant PHY03-00710.
\end{acknowledgements}

\appendix

\section{The Quasilocal Expansion}
\label{app_ql_exp}

In this appendix we provide the quasilocal expansions of the relevant quantities used in regulating the direct part of the retarded fields $A_\mu$ and $h_{\mu\nu}$ in Sections \ref{em} and \ref{grav}. The relations we start with have been derived in Appendix A of \cite{GH1} to which the reader is referred to for more details. The quasilocal expansion of Synge's world function is
\begin{eqnarray}
	\sigma^\alpha (z^\alpha, z^{\alpha^\prime}) &=& - s \, u^\alpha - \frac{s^2}{2!} \, a^\alpha - \frac{s^3}{6!} \, \frac{D a^\alpha }{ d\tau } - \ldots \non \\
	&=& - \sum_{n=1} ^\infty \frac{s^n}{n!} \, \left(\frac{D}{d\tau}\right)^{n-1} \, u^\alpha(\tau) ~.
	\label{sigma_vector}
\end{eqnarray}
From the identity $2\sigma = \sigma_\alpha \sigma^\alpha$ one can show that
\begin{eqnarray}
	\sigma = -\frac{ s^2}{2} - \frac{s^4 a^2}{24} + O (s^5)
	\label{sigma}
\end{eqnarray}
where $a^2=a_\alpha a^\alpha$ \footnote{A misprint in \cite{GH1} misses the $a^2$ factor in their Eqs. (A9) and (A10).}.


We can apply these results to compute the quasilocal expansion of the quantities appearing in the direct part of the retarded fields $A_\mu ^{ret}$ and $h_{\mu\nu} ^{ret}$. Following \cite{Poisson} the expansion of these quantities around $\sigma, \sigma^\alpha =0$ at $x^\alpha = z^\alpha (\tau)$ is
\begin{eqnarray}
	   && \!\!\! \Delta^{1/2} (z^\alpha, z^{\alpha^\prime} ) \nonumber \\
	   && {\hskip0.25in} = 1 + \frac{1}{12} \, R_{\alpha \beta} \, \sigma^\alpha \, \sigma^\beta - \frac{1}{24} \, R_{\alpha \beta ; \gamma} \, \sigma^\alpha \, \sigma^\beta \, \sigma^\gamma + \ldots ~, \nonumber \\
	   && \!\!\!  \nabla_{\nu} \Delta^{1/2} (z^\alpha, z^{\alpha^\prime} ) \nonumber \\
	   && {\hskip0.25in} =  \frac{1}{6} \ R_{\nu \alpha} \, \sigma ^\alpha - \frac{1}{24} \left( 2 R_{\nu \alpha ; \beta}  - R_{\alpha \beta; \nu}  \right) \, \sigma^\alpha \, \sigma^\beta  + \ldots ~. \nonumber \\
	   \label{vanVleck}
\end{eqnarray}
Invoking the quasilocal expansion $s = \tau^\prime - \tau$ and using (\ref{sigma_vector}) and (\ref{sigma}) results in the following 
\begin{eqnarray}
	&& \Delta^{1/2} (z^\alpha, z^{\alpha^\prime} ) = 1 + \frac{s^2 }{12} \, R_{\alpha \beta} u^\alpha u^\beta + O(s^3) ~, \nonumber \\
	&& \nabla_\nu \Delta^{1/2} (z^\alpha, z^{\alpha^\prime} ) = -\frac{s}{6} \, R_{\nu \alpha} u^\alpha + O(s^2) ~, \nonumber \\
	&& \delta_\Lambda (\sigma ) = \sqrt{ \frac{8}{\pi} } \, \Lambda^2 e^{-\Lambda^4 s^4/2} + O(s^6) ~,  \nonumber \\
	&& \nabla_\nu \delta_\Lambda (\sigma) = \sigma_\nu \left( \frac{\partial \sigma}{\partial s} \right) ^{-1} \frac{\partial \delta_\Lambda}{\partial s} \nonumber \\
	&& {\hskip0.1in} = \left( u_\nu + \frac{s}{2} \, a_\nu + \frac{s^2}{6} \, w_\mu ^{~\nu} \frac{D a_\nu}{d\tau} \right)  \frac{\partial \delta_\Lambda}{\partial s} + O(s^3) \nonumber \\
\end{eqnarray}
where $w_\mu ^{~\nu} = g_\mu ^{~\nu} + u_\mu u^\nu$ projects vectors onto a direction orthogonal to the 4-velocity $u^\mu$.

Lastly, we will need to determine the quasilocal expansion of $g_{\alpha \beta^\prime} u^{\beta^\prime}$ and $g_{\alpha \gamma^\prime ; \beta} u^{\gamma^\prime}$ when calculating the regularized self-force. These kinds of terms appear in the direct part of the retarded fields and result from integrating the direct part of the retarded Green's function with the current density. From \cite{Poisson} one can deduce the expansions around $\sigma, \sigma^\alpha = 0$ at $x^\alpha = z^\alpha (\tau)$ to be
\begin{eqnarray}
	&& g_{\alpha \beta^\prime} u^{\beta^\prime} = - \sigma_{\alpha \beta^\prime } u^{\beta^\prime} - \frac{1}{6} \, \sigma_{\beta^\prime} ^{~\beta} R_{\alpha \gamma \beta \delta} \sigma^\gamma \sigma^\delta u^{\beta^\prime} + \cdots ~, \nonumber \\
	&& g_{\alpha \gamma^\prime ; \beta} u^{\gamma^\prime} = \frac{1}{2} u^{\gamma^\prime} g_{\gamma^\prime} ^{~\epsilon} R_{\alpha \epsilon \beta \delta } \sigma^\delta + \cdots ~.
\end{eqnarray}
Recalling that $\sigma_{\alpha \beta^\prime } u^{\beta^\prime} = D \sigma_\alpha / d\tau^\prime$ and using the quasilocal expansions of $\sigma_\alpha$ and $\sigma$ from (\ref{sigma_vector}) and (\ref{sigma}), respectively, one can show that
\begin{eqnarray}
	&& g_{\alpha \beta^\prime} u^{\beta^\prime} = u_\alpha + s a_\alpha + \frac{s^2}{2} \, \frac{D a_\alpha}{ d\tau} + O (s^3) ~, \nonumber \\
	&& g_{\alpha \gamma^\prime ; \beta} u^{\gamma^\prime} = - \frac{s}{2} \, R_{\alpha \gamma \beta \delta}  u^\gamma u^\delta + O(s^2) ~.
\end{eqnarray}

\section{Time dependent coefficient in regulated dynamics}
\label{coefficients}

The $r$-dependent coefficients ($r=\tau- \tau_i$) appearing in the equations for the regulated semiclassical dynamics are
\begin{eqnarray}
    c_{(n)} (r) &=& - \int _{\tau_i} ^\tau d\tau^\prime \, \frac{s^n}{n!} \, \delta_\Lambda (s) \non \\ 
    &=& ( -1)^{n+1} \frac{ 2^{(n-1)/4} }{ \pi^{1/2} \, n! } \, \Lambda^{1-n} \, \gamma \left( \frac{1+n}{4} , \frac{r^4 \Lambda^4}{ 2} \right) \non \\ 
\end{eqnarray}
\begin{eqnarray}
    g_{(n)} (r) &=& \int _{\tau_i} ^\tau d\tau^\prime \, \frac{s^n}{n!} \, \frac{\partial \delta_\Lambda (s) }{ \partial s} \non \\ 
    &=& (-1)^n \: \frac{ 2^{(n+6)/4} }{ \pi ^{1/2} \: n! } \: \Lambda^{2-n} \, \gamma \left( 1 + \frac{n}{4} , \frac{r^4 \Lambda^4}{ 2} \right) \non \\
\end{eqnarray}
where $\gamma(a,b) = \Gamma(a) - \Gamma(a,b)$ and $\Gamma(a,b)$ is the incomplete gamma function. As $\Lambda$ or $r$ goes to infinity $\gamma(a,b) \rightarrow \Gamma (a)$. Since $c_{(n)}$ and $g_{(n)}$ are even functions of the elapsed time then we obtain the same values also when the initial time $\tau_i$ is in the remote past. In either limits both $c_{(1)}(r)$ and $g_{(2)}(r)$ approach $1$ on a time scale $r \sim 1/\Lambda$.

\section{Tensor coefficients in Linearized stochastic dynamics}
\label{stoch_coeff}

To determine the $\bar{z}$-dependent tensor coefficients appearing in (\ref{em_stoch_ALD}) and (\ref{grav_stoch_ALD}) we will need to make use of the following relations. Expanding the world line coordinates as $z^\mu = \bar{z}^\mu + \tilde{z}^\mu$ implies that the velocity is $u^\mu = \bar{u}^\mu + \tilde{u}^\mu$ and the acceleration
\begin{eqnarray}
	a^\mu &=& \frac{D u^\mu}{d\tau} = \bar{a}^\mu + \ddot{\tilde{z}}^\mu + 2 \bar{\Gamma} ^\mu_{\alpha \beta} \bar{u}^\alpha \dot{\tilde{z}}^\beta + \tilde{z}^\gamma \bar{\Gamma}^\mu _{\alpha \beta, \gamma} \bar{u}^\alpha \bar{u}^\beta \non \\
	&& + O(\tilde{z}^2)
\end{eqnarray}
An overbar indicates that the function is evaluated at the mean world line coordinate $\bar{z}^\mu$. One can also derive an expression for $Da^\mu/d\tau$ but we do not give it here. The indices are to be raised and lowered with the metric 
\begin{eqnarray}
	g_{\mu\nu} = \bar{g}_{\mu\nu} + \bar{g}_{\mu\nu, \gamma} \tilde{z}^\gamma
\end{eqnarray}
and the inverse metric is 
\begin{eqnarray}
	g^{\mu\nu} = \bar{g}^{\mu\nu} - \bar{g}^{\mu \alpha} \bar{g}^{\nu\beta} \bar{g}_{\alpha \beta, \gamma} \tilde{z}^\gamma
\end{eqnarray}
Using these relations we can determine the tensor coefficients $m_{\mu\nu} \ldots r_{\mu\nu}$ appearing in the equations of motion for the world line fluctuations $\tilde{z}^\mu$.

The $\bar{z}$-dependent tensor coefficients appearing in (\ref{em_stoch_ALD}) for the {\it electromagnetic} case are
\begin{widetext}
\begin{eqnarray}
	&& {\hskip-0.5in} m_{\mu\nu} [\bar{z}] = m_{ph} (r) \bar{g}_{\mu\nu} -2 e^2 g_{(2)}(r) \bar{w}_{\mu\alpha} \bar{\Gamma}^\alpha _{\beta \nu} \bar{u}^\beta \\
	&& {\hskip-0.5in} \gamma_{\mu\nu} [\bar{z}] = 2 m_{ph} (r) \bar{g}_{\mu\alpha} \bar{\Gamma}^\alpha _{\beta \nu} \bar{u}^\beta - \frac{2 e^2}{3} \, g_{(2)}(r) \bar{w}_{\mu\alpha} \left[ 2 \bar{\Gamma}^\alpha _{\beta \nu, \gamma} \bar{u}^\beta \bar{u}^\gamma + 2 \bar{\Gamma} ^\alpha _{\beta \nu} \dot{\bar{u}}^\beta + \bar{\Gamma}^\alpha _{\beta \gamma, \nu} \bar{u}^\beta \bar{u}^\gamma + \bar{\Gamma}^\alpha _{\beta \nu} \bar{a}^\beta + 2 \bar{\Gamma} ^\alpha _{\beta \gamma} \bar{\Gamma}^\beta _{\delta \nu} \bar{u}^\gamma \bar{u}^\delta \right] \non \\
	&& - \frac{4 e^2}{3} \, g_{(2)} (r) \bar{u}_{( \mu} \bar{g}^{\beta )} _{~\, \nu} \frac{D \bar{a}_\beta}{d\tau} - \frac{e^2}{6} \left( 3- c_{(1)}(r) \right) \left[ \bar{w}_\mu ^{~\alpha} \bar{R}_{\alpha \nu} + 2 \bar{u}_{(\mu} \bar{g}^{\alpha )} _{~\, \nu} \bar{R}_{\alpha \beta} \bar{u}^\beta \right] - 2 e \bar{g}_\mu ^{~[\alpha} \bar{g} ^{\beta ] }_{~\, \nu} \bar{A} ^{tail}_{\alpha \beta}  \\
	&& {\hskip-0.5in} \kappa_{\mu\nu} [\bar{z}] = m_{ph} (r) \left[ \bar{g}_{\mu\alpha} \bar{\Gamma}^\alpha _{\beta \gamma, \nu} \bar{u} ^\beta \bar{u}^\gamma + \bar{a}^\alpha \bar{g}_{\mu \alpha , \nu} \right] - \frac{e^2}{6} \left( 3- c_{(1)}(r) \right) \bar{w}_\mu ^{~\alpha} \bar{R} _{\alpha \beta , \nu} \bar{u}^\beta - e \bar{w}_\mu ^{~\alpha \beta } \bar{A}_{\alpha \beta, \nu} ^{tail} \non \\
	&& - \frac{2e^2}{3} \, g_{(2)}(r) \bar{w}_{\mu \alpha} \left[ \bar{g}^{\alpha \beta} \bar{g}_{\beta \gamma, \nu} \frac{D \bar{a}^\gamma}{ d\tau} + \bar{\Gamma}^\alpha _{\beta \gamma, \delta \nu} \bar{u}^\beta \bar{u}^\gamma \bar{u}^\delta + 2 \bar{\Gamma}^\alpha _{\beta \gamma, \nu} \dot{\bar{u}}^\beta \bar{u}^\gamma + \bar{\Gamma}^\alpha _{\beta \gamma} \bar{\Gamma}^\beta _{\delta \epsilon , \nu} \bar{u}^\gamma \bar{u}^\delta \bar{u}^\epsilon + \bar{\Gamma}^\alpha _{\beta \gamma, \nu} \bar{u}^\beta \bar{a}^\gamma \right]  \\
	&& {\hskip-0.5in} r_{\mu\nu} [\bar{z}] = \frac{2e^2}{3} \, g_{(2)}(r) \bar{w}_{\mu\nu}
\end{eqnarray}
where the convention is used that $T_{[\alpha \beta]} = ( T_{\alpha \beta} - T_{\beta \alpha}) /2$.
In flat spacetime and in Cartesian coordinates these become much simplified to
\begin{eqnarray}
	m_{\mu\nu} [\bar{z}] = m_{ph} (r) \eta_{\mu\nu} ~~&,& ~~ \gamma_{\mu\nu} [\bar{z}] = -\frac{4e^2}{3} \, g_{(2)}(r) \bar{u}_{(\mu} \dot{\bar{a}}_{\nu)} \\
	\kappa_{\mu\nu} [\bar{z}] = 0 ~~&,&~~ r_{\mu\nu} [\bar{z}] = \frac{2e^2}{3} \, g_{(2)}(r) \bar{w}_{\mu\nu}
\end{eqnarray}
where the symmetrization convention used is $T_{(\alpha \beta)} = ( T_{\alpha \beta} + T_{\beta \alpha}) /2$.

The $\bar{z}$-dependent tensor coefficients appearing in (\ref{grav_stoch_ALD}) for the {\it gravitational} case are
\begin{eqnarray}
	&& m_{\mu\nu}[\bar{z}] = m_o \bar{g}_{\mu\nu} \\
	&& \gamma_{\mu\nu} [\bar{z}]  = 2 m_0 \bar{g}_{\mu \alpha} \bar{\Gamma}^\alpha _{\beta \nu} \bar{u}^\beta + m_o \bar{h}_{\alpha \beta \gamma}^{tail} \left[ 2 \bar{u}_{(\mu } \bar{u}^\alpha \bar{u}^\beta \bar{g} ^{\gamma) } _{~\, \nu} + \bar{g}_\mu ^{~(\alpha} \bar{u}^{\beta)} \bar{g} ^\gamma _{~\nu} - \bar{g}_\mu ^{~(\alpha} \bar{u} ^{| \beta |} \bar{g}^{\gamma)} _{~\, \nu} + \bar{g}_\mu ^{~(\alpha} \bar{g}^{\beta)} _{~\nu} \bar{u}^\gamma \right] \\
	&& \kappa_{\mu\nu} [\bar{z}] = m_o \bar{a}^\alpha \bar{g}_{\mu \alpha, \nu} + m_o \bar{g}_{\mu\alpha} \bar{\Gamma}^\alpha _{\beta \gamma, \nu} \bar{u}^\beta \bar{u}^\gamma - m_o \bar{w}_\mu ^{~\alpha \beta \gamma} \bar{h}_{\alpha \beta \gamma, \nu} ^{tail}
\end{eqnarray}
In flat spacetime, $\gamma_{\mu\nu} = \kappa_{\mu\nu} = 0$ and (\ref{grav_stoch_ALD}) is simply $m_{o} \ddot{\tilde{z}}_\mu = \eta_\mu (\bar{z})$.

We also include corrected expressions for the tensor coefficients that appear in (5.7) of \cite{GH1}. Using their notation for the tensor coefficients we have
\begin{eqnarray}
	&& {\hskip-0.25in} M_{\mu\nu} [\bar{z}] = \left[ m^{ph} _\phi (r) -e^2 \int_{\tau_i} ^\tau d\tau^\prime \bar{V} \right] \bar{g}_{\mu\nu} - e^2 g_{(2)}(r) \bar{w}_{\mu\alpha} \bar{\Gamma}^\alpha _{\beta \nu} \bar{u}^\beta \\
	&& {\hskip-0.25in} \Gamma_{\mu\nu} [\bar{z}] = 2 \left[ m^{ph} _\phi (r) -e^2 \int_{\tau_i} ^\tau d\tau^\prime \bar{V} \right]  \bar{g}_{\mu\alpha} \bar{\Gamma}^\alpha _{\beta \nu} \bar{u}^\beta - \frac{2 e^2}{3} \, g_{(2)} (r) \bar{u}_{( \mu} \bar{g}^{\beta )} _{~\, \nu} \frac{D \bar{a}_\beta}{d\tau} - \frac{e^2}{6} c_{(1)}(r) \left[ \bar{w}_\mu ^{~\alpha} \bar{R}_{\alpha \nu} + 2 \bar{u}_{(\mu} \bar{g}^{\alpha )} _{~\, \nu} \bar{R}_{\alpha \beta} \bar{u}^\beta \right] \non \\
	&&  - \frac{e^2}{3} \, g_{(2)}(r) \bar{w}_{\mu\alpha} \left[ 2 \bar{\Gamma}^\alpha _{\beta \nu, \gamma} \bar{u}^\beta \bar{u}^\gamma + 2 \bar{\Gamma} ^\alpha _{\beta \nu} \dot{\bar{u}}^\beta + \bar{\Gamma}^\alpha _{\beta \gamma, \nu} \bar{u}^\beta \bar{u}^\gamma + \bar{\Gamma}^\alpha _{\beta \nu} \bar{a}^\beta + 2 \bar{\Gamma} ^\alpha _{\beta \gamma} \bar{\Gamma}^\beta _{\delta \nu} \bar{u}^\gamma \bar{u}^\delta \right]  - 2 e \bar{u}_{ (\mu} \bar{g} ^{\alpha ) }_{~\, \nu} \bar{\phi} ^{tail}_{\alpha}  \\
	&& {\hskip-0.25in} K_{\mu\nu} [\bar{z}] = \left[ m^{ph} _\phi (r) -e^2 \int_{\tau_i} ^\tau d\tau^\prime \bar{V} \right] \left[ \bar{g}_{\mu\alpha} \bar{\Gamma}^\alpha _{\beta \gamma, \nu} \bar{u} ^\beta \bar{u}^\gamma + \bar{a}^\alpha \bar{g}_{\mu \alpha , \nu} \right] - \frac{e^2}{6} \, c_{(1)}(r) \bar{w}_\mu ^{~\alpha} \bar{R} _{\alpha \beta , \nu} \bar{u}^\beta \non \\
	&& - \frac{e^2}{3} \, g_{(2)}(r) \bar{w}_{\mu \alpha} \left[ \bar{g}^{\alpha \beta} \bar{g}_{\beta \gamma, \nu} \frac{D \bar{a}^\gamma}{ d\tau} + \bar{\Gamma}^\alpha _{\beta \gamma, \delta \nu} \bar{u}^\beta \bar{u}^\gamma \bar{u}^\delta + 2 \bar{\Gamma}^\alpha _{\beta \gamma, \nu} \dot{\bar{u}}^\beta \bar{u}^\gamma + \bar{\Gamma}^\alpha _{\beta \gamma} \bar{\Gamma}^\beta _{\delta \epsilon , \nu} \bar{u}^\gamma \bar{u}^\delta \bar{u}^\epsilon + \bar{\Gamma}^\alpha _{\beta \gamma, \nu} \bar{u}^\beta \bar{a}^\gamma \right]  \\
	&& {\hskip-0.25in} r_{\mu\nu} [\bar{z}] = \frac{e^2}{3} \, g_{(2)}(r) \bar{w}_{\mu\nu}
\end{eqnarray}
where the physical mass of the scalar charge is $m^{ph} _\phi (r) = m_o - e^2 g_{(1)}(r) /2 +e^2 c_{(0}(r)$ and 
\begin{eqnarray}
	M_{\mu\nu} [\bar{z}] \ddot{\tilde{z}}^\nu + \Gamma_{\mu\nu} [\bar{z}] \dot{\tilde{z}}^\nu + K_{\mu\nu} [\bar{z}] \tilde{z}^\nu - e^2 \int _{\tau_i} ^\tau d\tau^\prime \left( \tilde{z}^\alpha \vec{w}_\mu [\bar{z}] \bar{V}_{;\alpha} + \tilde{z}^{\alpha^\prime} \vec{w}_\mu [\bar{z}] \bar{V}_{;\alpha^\prime} \right) = r_{\mu\nu} [\bar{z} ] \dddot{\tilde{z}}^\nu + O(\Lambda^{-1})
\end{eqnarray}
The operator $\vec{w}_\mu ^{~\alpha} [\bar{z}] = \bar{a}_\mu + \bar{w}_\mu ^{~\alpha} \nabla_\alpha$ is independent of $\tau^\prime$.
\\
\end{widetext}


\begin{thebibliography}{99}
\bibitem{footnote_1} In general, for systems with a nonlinear coupling to the environment or for non-Gaussian initial states of the environment the interpretation of $P_\xi$ as a probability distribution is not always possible. It may take on negative values in which case $P_\xi$ should be interpreted as a pseudo-probability distribution in a similar vein as the Wigner function \cite{Calzetta_Roura_Verdaguer}.
\end{thebibliography}
\end{document}